\documentclass[]{navercloud}

\title{\textsc{Sommelier}: Scalable Open Multi-turn Audio Pre-processing for Full-duplex Speech Language Models}

\author[1,2,\clubsuit]{Kyudan Jung}
\author[1,2,\clubsuit]{Jihwan Kim}
\author[2]{Soyoon Kim}
\author[2]{Jeonghoon Kim}
\author[1\dagger]{Jaegul Choo}
\author[1,2\dagger]{Cheonbok Park}

\affiliation{$^1$KAIST AI, $^2$NAVER Cloud}

\contribution[\dagger]{Corresponding authors}

\usepackage{etoolbox}
\usepackage{xspace}
\usepackage{adjustbox}
\usepackage[table,dvipsnames]{xcolor}  % in preamble
\usepackage{amssymb}
\usepackage{tabularx}
\usepackage{longtable}
\usepackage{booktabs}
\usepackage{threeparttable}
\usepackage{array}
\usepackage{pifont}
\usepackage{multirow}
\usepackage{wrapfig}
\usepackage{mdframed}
\usepackage{microtype}
\definecolor{goodred}{HTML}{E45A92}
\usepackage{times}
\usepackage{latexsym}
\usepackage{booktabs}
\usepackage{siunitx} % optional (숫자 정렬)
\usepackage{booktabs}
\usepackage{makecell}
\usepackage{multirow}

\usepackage{listings}
\lstdefinelanguage{json}{
  basicstyle=\ttfamily\scriptsize,
  numbers=left,
  numberstyle=\tiny,
  stepnumber=1,
  numbersep=6pt,
  showstringspaces=false,
  breaklines=true,
  frame=single,
  columns=fullflexible,
  morestring=[b]",
  stringstyle=\color{navergreen},
  keywords={true,false,null},
  keywordstyle=\color{goodred}\bfseries,
  literate=
   *{:}{{\color{black}:}}{1}
    {,}{{\color{black},}}{1}
    {\{}{{\color{black}\{}}{1}
    {\}}{{\color{black}\}}}{1}
    {[}{{\color{black}[}}{1}
    {]}{{\color{black}]}}{1}
}

\usepackage[T1]{fontenc}
\usepackage[utf8]{inputenc}

\usepackage{amsmath}
\usepackage{inconsolata}

\usepackage{graphicx}
\usepackage{float} % for [H] figure placement
\usepackage[table]{xcolor}
\newif\ifdraft  % set \draftfalse for submission
\drafttrue
\ifdraft

    \newcommand{\bsr}[1]{\textcolor{blue}{\sout{#1}}}
    \newcommand{\bsc}[1]{\textcolor{blue}{[BS: #1]}}
    \newcommand{\atj}[1]{\textcolor{purple}{(ATJ: #1)}}

\else

    \newcommand{\bsr}[1]{}
    \newcommand{\bsc}[1]{}
    \newcommand{\atj}[1]{}
\fi

\newcommand{\tablestyle}[2]{%
    \fontfamily{ptm}\selectfont%
    \let\itold\it%
    \def\it{\itold \fontfamily{ptm}\selectfont}%
    \setlength{\tabcolsep}{#1}\renewcommand{\arraystretch}{#2}\centering\kindatiny%
    \let\citeold\cite%
    \renewcommand{\cite}[1]{\normalfont\fontfamily{ptm}\selectfont\tiny\citeold{##1}}%
}
\newcommand{\bigtablestyle}[2]{%
    \fontfamily{ptm}\selectfont%
    \let\itold\it%
    \def\it{\itold \fontfamily{ptm}\selectfont}%
    \setlength{\tabcolsep}{#1}\renewcommand{\arraystretch}{#2}\centering\footnotesize%
    \let\citeold\cite%
    \renewcommand{\cite}[1]{\normalfont\fontfamily{ptm}\selectfont\footnotesize\citeold{##1}}%
}

\abstract{
As the paradigm of AI shifts from text-based LLMs to Speech Language Models (SLMs), there is a growing demand for full-duplex systems capable of real-time, natural human-computer interaction.
However, the development of such models is constrained by the scarcity of high-quality, multi-speaker conversational data, as existing large-scale resources are predominantly single-speaker or limited in volume.
Addressing the complex dynamics of natural dialogue, such as overlapping and back-channeling remains a challenge, with standard processing pipelines suffering from diarization errors and ASR hallucinations.
To bridge this gap, we present a robust and scalable open-source data processing pipeline designed for full-duplex model.
}

\correspondence{\email{cbok.park@navercorp.com}, \email{jchoo@kaist.ac.kr}}

\metadata[\optimisticfont Demo and Code]{\href{https://kyudan1.github.io/sommelier.github.io//}{\texttt{sommelier.github.io}}}

\makeatletter
\newcommand{\symfootnotetext}[2]{%
  \protected@xdef\@thefnmark{#1}%
  \@footnotetext{#2}%
}
\makeatother

\begin{document}

\maketitle
\symfootnotetext{$\clubsuit$}{This work was done during the residency program at NAVER Cloud.}
\begin{figure*}[t]
    \centering
    \includegraphics[width=1\linewidth, trim={0 0 0 0}, clip]{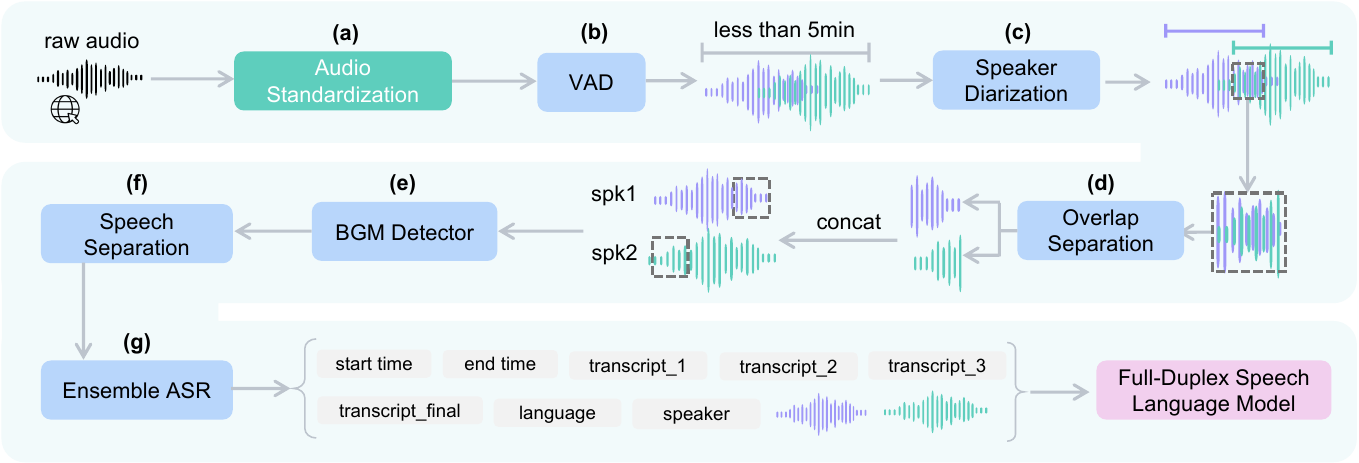}
    \caption{The overall pipeline of the \textsc{Sommelier} conversational audio pre-processing. Blue boxes denote neural model-based components, and a green box represent a algorithmic component.}
    \label{fig:pipeline}
\end{figure*}

\section{Introduction}

Recent advances in speech large language models (SLMs) have evolved from single short queries to multi-turn, open-ended conversations~\citep{xu2025qwen3omnitechnicalreport, goel2025audioflamingo3advancing}. Yet most systems still operate in disjoint user and assistant turns through a cascaded ASR and TTS pipeline~\citep{ye2025omnivinci}, which inherits latency, discards paralinguistic cues, and struggles with interruptions, overlapping speech, and backchanneling. Full-duplex system~\citep{wang2024fullduplexspeechdialoguescheme, roy2026personaplex} addresses these limitations by enabling the system to listen and speak simultaneously, supporting more fluid and human-like interaction.

Progress toward full-duplex SLMs has been facing bottlenecked by the lack of high-quality conversational data suitable for duplex training. While Moshi~\citep{defossez2024moshispeechtextfoundationmodel} leverages millions of hours of unsupervised audio for pre-training, these sources are largely single-stream and provide limited supervision for overlapping speech. Consequently, overlap robustness relies on relatively small high-fidelity conversational corpora such as Fisher~\citep{cieri2004fisher}, which is unlikely to meet the scale and diversity required for supervised fine-tuning (SFT)~\citep{xu2025qwen3omnitechnicalreport}.

Curating full-duplex training data from in-the-wild recordings is challenging because real conversations contain frequent overlaps, backchannels, and acoustic clutter, which amplify diarization and transcription errors~\citep{wang2024turn}. In addition, long-form audio typically includes non-conversational or irrelevant regions (e.g., music, noise, long silences), requiring careful filtering and normalization while preserving speaker structure and multi-turn context. Finally, processing web-scale audio demands high throughput to make large-scale curation feasible under practical compute budgets in the real industry.

To address these challenges, we propose an open, robust, and scalable speech pre-processing pipeline designed for full-duplex SLMs.
Our contributions are as follows:

\begin{itemize}
    \item \textsc{The f{}irst scalable pipeline for full-duplex SLMs:} We release a scalable pipeline for curating multi-turn conversational speech suitable for full-duplex training, helping alleviate the community-wide data scarcity.
    \item \textsc{High-f{}idelity overlap processing:} We provide a detailed processing strategy that handles overlaps via rigorous diarization analysis and reduces ASR hallucinations using paralleled model ensembling and n-gram filtering.
    \item \textsc{Proven eff{}icacy on a full-duplex model:} We validate our pipeline by fine-tuning the full-duplex model Moshi on \textsc{Sommelier}-processed speech and analyze practical data requirements for stable full-duplex training.
\end{itemize}

\section{Method}
In this section, we present \textsc{Sommelier}, a robust data processing pipeline designed to transform raw, in-the-wild conversational audio into high-quality training corpora for full-duplex Speech Language Models (SLMs).
Unlike traditional ASR pipelines that prioritize clean, non-overlapping speech, our design philosophy centers on preserving the chaotic yet rich dynamics of human dialogue, such as overlaps and backchannelings, while ensuring scalability for web-scale processing.
The overall architecture, illustrated in Figure~\ref{fig:pipeline}, is built as a modular framework where each component can be toggled or reconfigured, allowing researchers to adapt the trade-off between data purity and conversational authenticity.

\textsc{\textbf{Sommelier}} is designed to transform raw audio into clean, well-structured data while preserving the semantic context.
The process begins with standardization, bringing diverse audio formats into a unified representation. We then segment the audio based on silence detection, followed by a Voice Activity Detection (VAD) model~\citep{SileroVAD} that further partitions the content into chunks of less than five minutes, a practical constraint that prevents downstream models from running out of memory on lengthy recordings (\S~\ref{audio_standardization}).
Speaker diarization (\S~\ref{vad_speaker_diarization}) follows, identifying who speaks when. Guided by these speaker boundaries, we separate and restore overlapping speech regions (\S~\ref{handling-overlapping-speech}), with optional removal of background noise and music (\S~\ref{background_music_removal})depending on the use case. Finally, an ensemble of three Automatic Speech Recognition (ASR) models (\S~\ref{asr_ensemble}) generates text transcripts and captions, leveraging model diversity to improve robustness. Each module in the pipeline can be toggled on or off to suit specific requirements.

Rather than stripping away speech overlaps and backchannelings,  interruptions, and simultaneous speech that characterize real dialogue, we preserve them. This allows the duplex speech language model to learn not just what people say, but how conversations actually unfold.

\subsection{Audio Standardization}
\label{audio_standardization}
Since collected radio and podcast data vary in format and volume, we adopt the method of \citep{he2024emilia}. Using the \texttt{pydub}~\footnote{https://github.com/jiaaro/pydub} and \texttt{librosa}~\citep{mcfee2025librosa} libraries, we convert all audio to the standard format (16kHz, 16-bit, Mono) and perform loudness normalization to -20dBFS~\citep{he2024emilia} as illustrated in Figure~\ref{fig:pipeline}(a).

\subsection{VAD \& Speaker Diarization}
\label{vad_speaker_diarization}
To prevent out-of-memory issues with the diarization model, we split long audio files into units of less than 5 minutes as shown in Figure~\ref{fig:pipeline}(b).
To maintain conversational context, we use a VAD model to cut the audio at silence intervals.

For speaker diarization, as shown in Figure~\ref{fig:pipeline}(c), instead of the commonly used pyannote \texttt{speaker-diarization-3.1} model \citep{pyannote1}, we adopted \textit{Sortformer}~\citep{parksortformer} from NVIDIA.
Section \ref{exp_diarization} presents a performance comparison between the two models, demonstrating Sortformer's superiority in robustly capturing very short utterances such as backchannelings.

\subsection{Handling Overlapping Speech}
\label{handling-overlapping-speech}
Conversational audio  features frequent turn changes and short utterances~\citep{wang2025companioncastmultiagentconversationalai}.
\begin{figure}
    \centering
    \includegraphics[width=0.6\linewidth, trim={0 0 0 5}, clip]{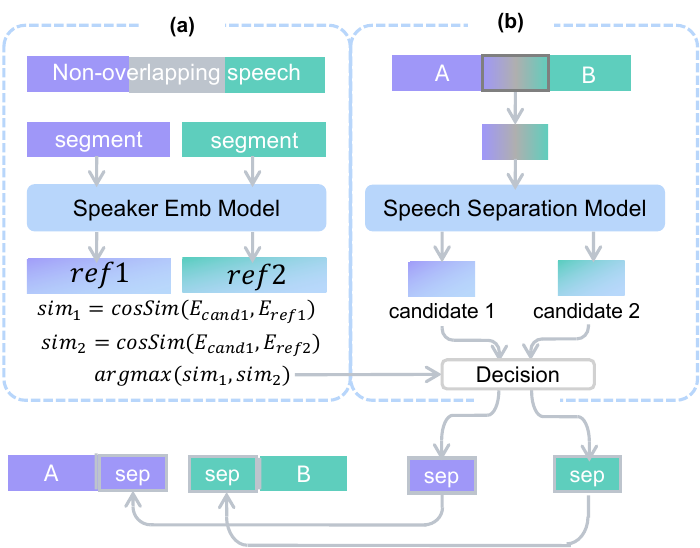}
    \caption{Illustration of the speech overlap separation process. \textbf{(a)} The process of calculating similarity to distinguish speaker identities using arbitrary independent speaker segments. \textbf{(b)} Separating overlapped regions and making identity decisions for candidates based on the similarity calculated in (a). Finally, the separated segments are concatenated with the original segments.}
    \label{fig:overlap_method}
\end{figure}
To systematically handle massive-scale industrial speech data, we categorized overlapping scenarios into four distinct cases, as shown in Figure~\ref{fig:overlap}.
\textit{Case 1} segments based on the overlap, yielding non-overlapping segments but losing full utterance information.
\textit{Cases 2} and \textit{Case 3} assign the overlapping speech to one side, which risks ASR errors where utterances mix or transcripts fail.
\textit{Case 4} allows both segments to contain the overlap based on speaker identity, preserving full information despite sharing the ASR issues of \textit{Cases 2} and \textit{Case 3}.

We selected \textit{Case 4} as our baseline, incorporating a module that performs two-speaker separation~\citep{shin2024separate} on the overlapped intervals.
We find that inputting only the duplicated part into separation model works better than using the entire segment.

Before separating overlapped speech, we extract non-overlapping parts longer than 2 seconds to generate embedding tuples $(e_{ref_1}, e_{ref_2})=(\mathcal{M}_{emb}(a_0), \mathcal{M}_{emb}(a_1))$, where $a$ denotes the audio segment and $\mathcal{M}$ represents the model. This process uses the speaker embedding model $\mathcal{M}_{emb}$ as shown in Figure~\ref{fig:overlap_method}(a).
In parallel, the overlapped audio $a_{overlap}$ is fed into the speech separation model $\mathcal{M}_{sep}$ to produce candidates $a_{cand1}$ and $a_{cand2}$ as shown in Figure~\ref{fig:overlap_method}(b). To identify the speakers, we calculate the cosine similarity scores $S_1 = \text{sim}(\mathcal{M}_{emb}(a_{cand1}), e_{ref_1})$ and $S_2 = \text{sim}(\mathcal{M}_{emb}(a_{cand1}), e_{ref_2})$.
The candidate with the higher similarity is assigned to the corresponding speaker ($a_i$), while the other candidate corresponds to the remaining speaker.
Finally, we concatenate the non-overlapping parts with the separated segments to create single-speaker segments as shown in Figure~\ref{fig:pipeline}(d).

\subsection{Background Music Removal}
\label{background_music_removal}
In addition to multi-speaker overlaps, the diverse nature of industrial audio sources introduces another challenge. Audio from radio broadcasts or dramas contains background music (BGM), which may be undesirable for training speech language models. We employ PANNs~\citep{kong2020panns} (Pre-trained Audio Neural Networks) to estimate the probability of background music presence in each segment. If the probability exceeds a threshold of 0.3, we apply the Demucs~\citep{rouard2022hybrid,defossez2021hybrid} model to extract the vocal track. Since music removal can degrade speech quality, we selectively apply it only to segments identified by PANNs, minimizing unnecessary processing as shown in Figure~\ref{fig:pipeline}(e) and (f).

We find that feeding the entire audio context into Demucs yields substantially better separation performance than processing short segments in isolation. Therefore, we input full two-minute audio chunks into the model and subsequently extract only the required portions from the separated output.
We also considered SAM-Audio~\citep{shi2025samaudiosegmentaudio} for music removal but excluded it due to its high inference latency (RTF 0.73 on A100), which limits its scalability for large datasets.

\subsection{Ensemble-based ASR}
\label{asr_ensemble}
High-quality ASR is essential for constructing large-scale datasets, as it generates the text labels necessary for model training.
However, relying on a single model, even SOTA architectures like Whisper~\citep{radford2022robustspeechrecognitionlargescale}, poses significant risks.
These models are prone to hallucinations, particularly in silent or noisy segments, where they often generate repetitive or nonsensical text~\citep{whisperhallucination1, whisperhallucination2, whisperhallucination3}.
Such artifacts introduce noise into the training signal, causing the downstream model to mimic these pathological behaviors.
\begin{table*}[t]
\centering
\scriptsize
\resizebox{\textwidth}{!}{%
\begin{tabular}{l|cc|ccc|cc|ccc}
\toprule
\multirow{2}{*}{\opthead{Model}} & \multicolumn{2}{c|}{\opthead{Pause Handling}} & \multicolumn{3}{c|}{\opthead{Backchannel}} & \multicolumn{2}{c|}{\opthead{Smooth Turn Taking}} & \multicolumn{3}{c}{\opthead{User Interruption}} \\
\cmidrule(lr){2-3} \cmidrule(lr){4-6} \cmidrule(lr){7-8} \cmidrule(lr){9-11}
& \makecell[c]{Synthetic\\TOR $\downarrow$} & \makecell[c]{Candor\\TOR $\downarrow$}
& \makecell[c]{TOR\\$\downarrow$} & \makecell[c]{Freq\\$\uparrow$} & \makecell[c]{JSD\\$\downarrow$}
& \makecell[c]{Candor\\TOR $\uparrow$} & \makecell[c]{Latency\\$\downarrow$}
& \makecell[c]{TOR\\$\uparrow$} & \makecell[c]{GPT-4o\\$\uparrow$} & \makecell[c]{Latency\\$\downarrow$} \\
\midrule
\texttt{Moshi} & 0.985 & 0.980 & 1.000 & 0.001 & 0.957 & 0.941 & 0.265 & 1.000 & 0.765 & 0.257 \\
\rowcolor{naverbg}\textbf{\texttt{Moshi}+\textsc{Sommelier}} & 1.000 & 1.000 & \textbf{0.291} & \textbf{0.052} & \textbf{0.630} & \textbf{1.000} & 0.344 & 0.858 & \textbf{3.684} & 1.065 \\
\bottomrule
\end{tabular}}
\caption{Full-Duplex-Bench 1.0 results for base Moshi and Moshi fine-tuned on 83 hours of \textsc{Sommelier}-processed data. Arrows indicate whether higher ($\uparrow$) or lower ($\downarrow$) values are better.}
\label{tab:finetuning}
\end{table*}
To mitigate this, we employ a Recognizer Output Voting Error Reduction (ROVER)~\citep{rover} ensemble strategy combining outputs from three distinct SOTA models as shown in Figure~\ref{fig:pipeline}(g).
We align transcripts at the word level and apply a prioritized majority voting scheme: a word is accepted if predicted by at least two models; otherwise, we default to the prediction of our primary backbone, Whisper, to maintain consistency.
Residual hallucinations are further pruned using a \texttt{RepetitionFilter} that discards samples with excessive n-gram ($n=15$) repetitions (count $\ge 5$)~\citep{udandarao2025datacentriclessonsimprovespeechlanguage}.
Concurrently, we extract word-level timestamps via Whisper.
Precision in temporal alignment is critical for modern streaming speech language models (as detailed in Section~\ref{exp_finetuning}), which typically require strict synchronization between audio and text tokens.

\section{Experiments}
\label{experiments}
In this section, we validate the individual components of our proposed pipeline. First, we examine the practical utility of our approach by fine-tuning Moshi using the dataset preprocesed by our proposed pipeline (\S~\ref{exp_finetuning}). We then quantitatively evaluate the diarization accuracy (\S~\ref{exp_diarization}), the audio quality following overlap separation (\S~\ref{exp_overlap}), and the accuracy of the ensemble-based ASR (\S~\ref{exp_asr}). Furthermore, to provide a comprehensive analysis, we discuss the pipeline's latency (\S~\ref{exp_latency}).

\subsection{Effectiveness of \textsc{Sommelier}-Processed Data for Full-Duplex Models}
\label{exp_finetuning}
To validate the effectiveness of our proposed \textsc{Sommelier} pipeline, we examine whether training a full-duplex model on the data processed by this pipeline yields performance improvements. To this end, we performed LoRA fine-tuning on \texttt{moshiko-pytorch-bf16}~\citep{defossez2024moshispeechtextfoundationmodel} and evaluated its duplex performance using the Full-Duplex-Bench~\citep{fullduplexbench1,fullduplexbench2}.

\begin{table}[t]
\centering
\resizebox{0.48\textwidth}{!}{%
\begin{tabular}{lcccc}
\toprule
\opthead{Model} &
\opthead{DER (\%)} &
\opthead{JER (\%)} &

\makecell[c]{\opthead{DER}\\\opthead{($\le$1.0s, \%)}} &
\makecell[c]{\opthead{DER}\\\opthead{(turn, \%)}}  \\
\midrule
pyannote3.1    & 8.40 & 17.68 & 20.21 & 0.051  \\
\rowcolor{naverbg}sortformer\_v1 & \textbf{7.16} & \textbf{14.69} & \textbf{16.87} & \textbf{0.006}  \\
\bottomrule
\end{tabular}%
}
\caption{Diarization model ablation on VoxConverse~\citep{voxconverse} (common subset, $\le$4 speakers). Lower is better for DER/JER and RTF.}
\label{tab:diar_ablation_voxconverse}
\end{table}

\textsc{Dataset.} We find that long turn-taking in the training data for Moshi, specifically when a single speaker holds the floor for too long (more than a minute), leads to unstable loss reduction and degrades performance, causing the model to become unresponsive. Consequently, we selected segments from the \textsc{Sommelier}-processed data where each turn lasted no more than 10 seconds. We defined a valid region as a sequence of at least three consecutive turns, truncating the region if an utterance exceeding 10 seconds appeared. Additionally, we assigned only a single speaker to the left channel of the stereo training data. We confirmed that these selection criteria significantly impact the training dynamics.
The model configuration is detailed in Appendix~\ref{app:detail_of_finetuning}.

\textsc{Results.} As shown in Table~\ref{tab:finetuning}, evaluation on Full-Duplex-Bench 1.0 demonstrated performance improvements across Backchanneling, Smooth Turn-Taking, and User Interruption handling. Regarding Pause Handling, however, we observe that the model performs comparably to base Moshi, exhibiting similar limitations. We hypothesize that this stems from the Moshi architecture or the absence of prompt audio, as proposed in Personaplex~\citep{roy2026personaplex}. Regarding latency, the base model exhibited notably short latencies simply because it failed to engage in backchanneling or interruption handling, reflecting suboptimal behavior where the model continued speaking regardless of user input. In contrast, after fine-tuning with Sommelier-processed data, the increased latency can be interpreted positively, as it indicates that the model is now actively processing user input and preparing appropriate responses for backchannels and interruptions. Detailed descriptions of the benchmark metrics are provided in Appendix~\ref{app:detail_of_finetuning}.

\begin{table}[t]
\centering
\resizebox{0.6\textwidth}{!}{%
\begin{tabular}{cc|ccc|ccc|ccc}
\toprule
\multirow{2}{*}{\opthead{SIR}} & \multirow{2}{*}{\opthead{OVL}} & \multicolumn{3}{c|}{\opthead{WER (\%) ↓}} & \multicolumn{3}{c|}{\opthead{STOI ↑}} & \multicolumn{3}{c}{\opthead{UTMOS ↑}} \\
 & & Ori & Sep & Orc & Ori & Sep & Orc & Ori & Sep & Orc \\
\midrule
\multirow{3}{*}{0 dB} 
 & 0.2 & 10.5 & \cellcolor{naverbg}\textbf{6.1} & 4.8 & .961 & \cellcolor{naverbg}\textbf{.982} & 1.00 & 3.04 & \cellcolor{naverbg}\textbf{3.53} & 3.88 \\
 & 0.5 & 13.9 & \cellcolor{naverbg}\textbf{7.9} & 5.8 & .888 & \cellcolor{naverbg}\textbf{.969} & 1.00 & 2.27 & \cellcolor{naverbg}\textbf{3.32} & 3.87 \\
 & 1.0 & 48.9 & \cellcolor{naverbg}\textbf{15.6} & 5.3 & .778 & \cellcolor{naverbg}\textbf{.913} & 1.00 & 1.70 & \cellcolor{naverbg}\textbf{3.02} & 3.84 \\
\midrule
\multirow{3}{*}{5 dB} 
 & 0.2 & 11.3 & \cellcolor{naverbg}\textbf{7.6} & 5.3 & .961 & \cellcolor{naverbg}\textbf{.978} & 1.00 & 3.06 & \cellcolor{naverbg}\textbf{3.47} & 3.87 \\
 & 0.5 & 18.8 & \cellcolor{naverbg}\textbf{7.1} & 4.3 & .887 & \cellcolor{naverbg}\textbf{.971} & 1.00 & 2.34 & \cellcolor{naverbg}\textbf{3.39} & 3.91 \\
 & 1.0 & 52.5 & \cellcolor{naverbg}\textbf{9.1} & 4.0 & .761 & \cellcolor{naverbg}\textbf{.936} & 1.00 & 1.79 & \cellcolor{naverbg}\textbf{3.12} & 3.91 \\
\midrule
\multirow{3}{*}{10 dB} 
 & 0.2 & 12.6 & \cellcolor{naverbg}\textbf{7.0} & 5.6 & .961 & \cellcolor{naverbg}\textbf{.980} & 1.00 & 3.26 & \cellcolor{naverbg}\textbf{3.60} & 3.98 \\
 & 0.5 & 29.7 & \cellcolor{naverbg}\textbf{10.1} & 5.2 & .877 & \cellcolor{naverbg}\textbf{.956} & 1.00 & 2.58 & \cellcolor{naverbg}\textbf{3.21} & 3.86 \\
 & 1.0 & 51.0 & \cellcolor{naverbg}\textbf{13.8} & 4.8 & .754 & \cellcolor{naverbg}\textbf{.919} & 1.00 & 2.17 & \cellcolor{naverbg}\textbf{3.01} & 3.92 \\
\bottomrule
\end{tabular}%
}
\caption{Speech quality for separated overlapped speech across metrics for \textbf{Ori}ginal audio, source \textbf{Sep}arated, and \textbf{Or}a\textbf{c}le (pre-synthesis speech quality).}
\label{tab:summary}
\end{table}
\subsection{Diarization Model Choice}
\label{exp_diarization}
Using \texttt{Pyannote 3.1} has been widely regarded as the default for diarization models, a trend followed by recent works such as \citet{he2024emilia} . In this study, however, we compare the performance of Sortformer~\citep{parksortformer}, which is adopted in our pipeline, against the Pyannote 3.1~\citep{pyannote1, pyannote2} baseline.

\textsc{Metrics.}
We evaluate speaker diarization quality using DER (Diarization Error Rate) and JER (Jaccard Error Rate).
DER measures the fraction of speaker time that is incorrectly attributed, typically aggregating \textit{missed speech}, \textit{false alarm speech}, and \textit{speaker confusion} within a tolerance collar.
JER measures the average Jaccard distance between the reference and hypothesis speaker segments, and is known to be more sensitive to boundary quality and segmentation consistency.
To stress-test challenging regimes, we additionally compute \textit{DER on short-duration speech} by restricting evaluation to reference segments shorter than a threshold ($\le$0.5\,s and $\le$1.0\,s), and \textit{DER on turn-taking regions} by restricting evaluation to temporal windows around speaker change points (speaker alternations within a small gap).
All metrics are reported on the VoxConverse~\citep{voxconverse} common subset containing recordings with at most four speakers.

\textsc{Results and analysis.}
Table~\ref{tab:diar_ablation_voxconverse} demonstrates that Sortformer consistently outperforms the Pyannote~3.1 baseline across global metrics on the VoxConverse benchmark. More importantly, the gains are most pronounced in regimes critical for conversational modeling. Sortformer exhibits superior robustness in handling short utterances and rapid turn-taking, effectively reducing errors in brief interjections and speaker boundaries. These results confirm that Sortformer is better suited for processing highly interactive, overlapping dialogue than standard baselines.

\subsection{Speech Quality of Overlap Separation}
\label{exp_overlap}

Processing overlapped speech is a critical step in constructing training datasets for full-duplex conversational models~\citep{defossez2024moshispeechtextfoundationmodel}. This is because full-duplex training requires speech segments to overlap freely, as in natural human conversations, while maintaining source-separated audio streams for each speaker.

Given two speech segments $a_i$ and $a_j$ that are sequentially overlapped, where $a_i$ starts at $t_{start}$, $a_j$ ends at $t_{end}$, and the overlap occurs from $t_1$ to $t_2$, we evaluate speech quality for each diarized speaker's utterance interval: $[t_{start}, t_2]$ for \texttt{speaker1} and $[t_1, t_{end}]$ for \texttt{speaker2}.

\textsc{Dataset.} To simulate diverse real-world overlap conditions, we synthesized 900 samples of two speaker mixtures from the LibriSpeech~\citep{librispeech} test utterance by varying Signal-to-Interference Ratio (SIR $\in \{0, 5, 10\}$ dB) and overlap ratio ($\rho \in \{0.2, 0.5, 1.0\}$), forming nine different conditions. We also mix silence-trimmed sources to achieve the target overlap precisely.

\begin{table}[t]
\centering
\resizebox{0.45\textwidth}{!}{%
\begin{tabular}{llcc}
\toprule
\opthead{Dataset} & \opthead{Model} & \opthead{WER (\%)} & \opthead{Time (s)} \\
\midrule
% --- LibriSpeech Test Clean ---
% [-0.3em]을 추가하여 텍스트를 아래로 내림
\multirow{2}{*}[-0.3em]{\shortstack{LibriSpeech\\Test Clean}}
 & Whisper   & $3.63 \pm 9.37$ & 0.39 \\

 \cmidrule(l){2-4}
 & \cellcolor{naverbg} \textbf{MoE (Ours)} &\cellcolor{naverbg}  $2.04 \pm 6.50$ & \cellcolor{naverbg} 1.40 \\
\midrule
% --- LibriSpeech Test Other ---
\multirow{2}{*}[-0.3em]{\shortstack{LibriSpeech\\Test Other}}
 & Whisper   & $6.26 \pm 11.63$ & 0.35 \\

 \cmidrule(l){2-4}
 & \cellcolor{naverbg} \textbf{MoE (Ours)} & \cellcolor{naverbg} $3.92 \pm 8.92$ &\cellcolor{naverbg}  1.27 \\
\midrule

% --- TEDLIUM3 Test ---
\multirow{2}{*}[-0.3em]{\shortstack{TEDLIUM3\\Test}}
 & Whisper   & $12.19 \pm 12.31$ & 0.36 \\

 \cmidrule(l){2-4}

 &\cellcolor{naverbg} \textbf{MoE (Ours)} &\cellcolor{naverbg}  $10.66 \pm 11.73$ & \cellcolor{naverbg} 1.33 \\
\bottomrule
\end{tabular}%
}
\caption{Evaluation results on LibriSpeech (Clean/Other) and TEDLIUM3. Whisper refers \texttt{Whisper-large-v3} model.}
\label{tab:asr}
\end{table}

\textsc{Metrics.} Evaluation is conducted across three conditions: (1) \textit{Original}, which directly extracts time segments from the mixed signal, (2) \textbf{Sep}, which applies SepReformer~\citep{shin2024separate}-based separation with speaker identity matching (see Section~\ref{handling-overlapping-speech}), and (3) \textit{Oracle}, which uses clean source signals as an upper bound. Ground-truth diarization timestamps from data synthesis are used across all conditions to ensure fair evaluation of overlapped regions. We assess intelligibility using Word Error Rate (WER), acoustic quality using SI-SDR and STOI, and perceptual naturalness using UTMOS~\citep{saeki2022utmos}.

\textsc{Results and Analysis.}
The quantitative analysis presented in Table \ref{tab:summary} reveals that while variations in the Signal-to-Interference Ratio (SIR) have a limited impact on performance, the overlap ratio serves as the critical determinant of task difficulty. As the overlap ratio increases, the baseline model suffers significant degradation; however, our proposed method (Sep) consistently outperforms the baseline across all experimental conditions, demonstrating robust separation capabilities even in highly overlapped scenarios. Most notably, in terms of perceptual quality (UTMOS~\citep{saeki2022utmos}), the proposed method achieves scores closely approximating the Oracle upper bound. This result strongly suggests that our model not only improves intelligibility but also preserves speech naturalness effectively, thereby guaranteeing the generation of high-quality samples suitable for use as training data.
More detailed results and analysis are provided in Appendix~\ref{app:detail_overlap}.

\subsection{ASR Ensemble Performance}
\label{exp_asr}
We compare the performance of the  Whisper model against our proposed three-model ensemble method utilizing ROVER~\cite{rover}.

\textsc{Metrics.} To evaluate ASR performance, we measure the Word Error Rate (WER) using the LibriSpeech~\citep{librispeech} test dataset (clean and other splits) and the TEDLIUM3~\citep{ted_lium} test set. Since LibriSpeech \texttt{test-other} and TEDLIUM3 contain noisy conditions, these datasets allow us to assess the model's robustness against real-world scenarios.

\textsc{Results and Analysis.} The comparison results between the standalone Whisper Large v3 and the ensemble (Whisper + Canary + Parakeet \citep{sekoyan2025canary1bv2parakeettdt06bv3efficient}) are presented in Table~\ref{tab:asr}. In terms of WER, the proposed approach demonstrated a significant improvement of approximately 37\%, reducing the error rate from 6.26\% to 3.92\% compared to the single \texttt{Whisper-large-v3} baseline. This gap was particularly evident in noisy data, demonstrating improved recognition accuracy in segments containing low volume or BGM.

Regarding inference time, the ensemble approach required approximately three times longer than the baseline. This latency is primarily attributed to the inference speed of Canary, the slowest model among the three, rather than sequential execution. Additionally, the concurrent loading and inference of three models may introduce slight overhead.

Qualitatively, we observed hallucinations in the Whisper outputs, such as repetitive generation (e.g., \textit{``Yeah., Yeah., Yeah...''}). We confirmed that our method successfully corrected these errors by selecting the accurate transcript provided by Canary (e.g., \textit{``Yeah, big decision for Dan''}).

\subsection{Latency}
\label{exp_latency}
\begin{table}[t]
\centering
\resizebox{0.47\textwidth}{!}{%
\begin{tabular}{lcc}
\toprule
\opthead{Stage} & \opthead{Processing Time (s)} & \opthead{RTF} \\
\midrule
Audio Duration & 120.00 & -- \\
\midrule
VAD + Sortformer & 1.91 & 0.0159 \\
SepReformer Separation & 0.15 & 0.0013 \\
ASR ensemble & 13.91 & 0.1159 \\
FlowSE Denoising & 4.99 & 0.0416 \\
\midrule
\rowcolor{naverbg}
\textbf{Total} & \textbf{20.95} & \textbf{0.1746} \\
\bottomrule
\end{tabular}}
\caption{Processing time breakdown for the proposed pipeline on a 120-second audio sample.}
\label{tab:processing_time}
\end{table}
Data preprocessing is a computationally intensive task~\cite{he2024emilia,Dua_2025}. Thus, minimizing latency in this phase is critical. As shown in Table~\ref{tab:processing_time}, running a single process on an A100 (80GB) yields a total Real-Time Factor (RTF) of 0.1746. Excluding the optional FlowSE Denoising step further reduces the RTF to 0.133, with the primary bottleneck occurring in the ASR stage. Given the peak memory usage of 23GB, it is possible to allocate three concurrent processes on a single GPU, which effectively lowers the RTF to 0.0443 per GPU. Consequently, processing 10,000 hours of audio using eight A100 GPUs would take approximately 55 hours, demonstrating the practical feasibility of our approach.

\section{Conclusion} 
We presented \textsc{Sommelier}, the first scalable, open-source pipeline for full-duplex SLMs.
Our pipeline combines rigorous diarization, overlap handling, and ensemble-based ASR to improve overall transcript quality.
We validated the overall utility of the \textsc{Sommelier} pipeline by fine-tuning Moshi on \textsc{Sommelier}-processed speech. We release our pipeline to support reproducible industrial research and to accelerate progress toward natural, real-time human--AI interaction.

\section*{Limitations}
A limitation of our pipeline is its exclusive focus on processing speech data. While optimized for conversational dialogue, it does not explicitly account for non-speech acoustic events or general sound scenes, limiting its scope compared to omni-modal audio approaches. Although our overlap separation module effectively disentangles simultaneous speakers from single-stream recordings, the resulting audio fidelity is inevitably slightly inferior to datasets that are originally recorded with distinct, isolated channels (Oracle), as the artificial separation process may introduce minor acoustic artifacts.

\section*{Ethical Considerations}
We developed the \textsc{Sommelier} pipeline with a strict adherence to open-source compliance and intellectual property rights. All software components, libraries, and pre-trained models integrated into our framework are governed by commercially permissive licenses, primarily MIT and Creative Commons (CC), allowing for broad academic and industrial application without legal ambiguity.

Furthermore, the podcast audio samples featured on our project demonstration page were exclusively selected from sources explicitly released under CC licenses. We have rigorously verified the usage terms of these recordings to ensure that no copyrighted material is infringed upon and that the original creators' rights are respected.

Beyond licensing compliance, we acknowledge the broader implications of releasing tools for high-fidelity speech processing. While our goal is to advance full-duplex interaction, we recognize that high-quality conversational datasets can potentially be misused for non-consensual voice cloning or deepfake generation. We urge the research community to utilize this pipeline responsibly, ensuring that any private data processed is done so with appropriate consent and privacy safeguards in place.

\section*{Acknowledgments}
We would like to express our deepest gratitude to Taehong Moon.

\bibliographystyle{assets/plainnat}
\bibliography{paper}

\begin{thebibliography}{59}
\providecommand{\natexlab}[1]{#1}
\providecommand{\url}[1]{\texttt{#1}}
\expandafter\ifx\csname urlstyle\endcsname\relax
  \providecommand{\doi}[1]{doi: #1}\else
  \providecommand{\doi}{doi: \begingroup \urlstyle{rm}\Url}\fi

\bibitem[AI et~al.(2025)AI, Ma, Zou, Yan, Jin, Shen, Zheng, Wang, Xu, Yao, et~al.]{ai2025ming}
Inclusion AI, Bowen Ma, Cheng Zou, Canxiang Yan, Chunxiang Jin, Chunjie Shen, Dandan Zheng, Fudong Wang, Furong Xu, GuangMing Yao, et~al.
\newblock Ming-flash-omni: A sparse, unified architecture for multimodal perception and generation.
\newblock \emph{arXiv preprint arXiv:2510.24821}, 2025.

\bibitem[Bara{\'n}ski et~al.(2025)Bara{\'n}ski, Jasi{\'n}ski, Bartolewska, Kacprzak, Witkowski, and Kowalczyk]{whisperhallucination2}
Mateusz Bara{\'n}ski, Jan Jasi{\'n}ski, Julitta Bartolewska, Stanis{\l}aw Kacprzak, Marcin Witkowski, and Konrad Kowalczyk.
\newblock Investigation of whisper asr hallucinations induced by non-speech audio.
\newblock In \emph{ICASSP 2025-2025 IEEE International Conference on Acoustics, Speech and Signal Processing (ICASSP)}, pages 1--5. IEEE, 2025.

\bibitem[Borsos et~al.(2023)Borsos, Marinier, Vincent, Kharitonov, Pietquin, Sharifi, Roblek, Teboul, Grangier, Tagliasacchi, and Zeghidour]{borsos2023audiolmlanguagemodelingapproach}
Zalán Borsos, Raphaël Marinier, Damien Vincent, Eugene Kharitonov, Olivier Pietquin, Matt Sharifi, Dominik Roblek, Olivier Teboul, David Grangier, Marco Tagliasacchi, and Neil Zeghidour.
\newblock Audiolm: a language modeling approach to audio generation, 2023.
\newblock \url{https://arxiv.org/abs/2209.03143}.

\bibitem[Bredin(2023)]{pyannote1}
Hervé Bredin.
\newblock {pyannote.audio 2.1 speaker diarization pipeline: principle, benchmark, and recipe}.
\newblock In \emph{Proc. INTERSPEECH 2023}, 2023.

\bibitem[Chen et~al.(2021)Chen, Chai, Wang, Du, Zhang, Weng, Su, Povey, Trmal, Zhang, Jin, Khudanpur, Watanabe, Zhao, Zou, Li, Yao, Wang, You, and Yan]{gigaspeech}
Guoguo Chen, Shuzhou Chai, Guan-Bo Wang, Jiayu Du, Wei-Qiang Zhang, Chao Weng, Dan Su, Daniel Povey, Jan Trmal, Junbo Zhang, Mingjie Jin, Sanjeev Khudanpur, Shinji Watanabe, Shuaijiang Zhao, Wei Zou, Xiangang Li, Xuchen Yao, Yongqing Wang, Zhao You, and Zhiyong Yan.
\newblock Gigaspeech: An evolving, multi-domain asr corpus with 10,000 hours of transcribed audio.
\newblock In \emph{Interspeech 2021}. ISCA, 2021.
\newblock \doi{10.21437/interspeech.2021-1965}.
\newblock \url{http://dx.doi.org/10.21437/Interspeech.2021-1965}.

\bibitem[Chu et~al.(2023)Chu, Xu, Zhou, Yang, Zhang, Yan, Zhou, and Zhou]{chu2023qwenaudioadvancinguniversalaudio}
Yunfei Chu, Jin Xu, Xiaohuan Zhou, Qian Yang, Shiliang Zhang, Zhijie Yan, Chang Zhou, and Jingren Zhou.
\newblock Qwen-audio: Advancing universal audio understanding via unified large-scale audio-language models, 2023.
\newblock \url{https://arxiv.org/abs/2311.07919}.

\bibitem[Chu et~al.(2024)Chu, Xu, Yang, Wei, Wei, Guo, Leng, Lv, He, Lin, Zhou, and Zhou]{chu2024qwen2audiotechnicalreport}
Yunfei Chu, Jin Xu, Qian Yang, Haojie Wei, Xipin Wei, Zhifang Guo, Yichong Leng, Yuanjun Lv, Jinzheng He, Junyang Lin, Chang Zhou, and Jingren Zhou.
\newblock Qwen2-audio technical report, 2024.
\newblock \url{https://arxiv.org/abs/2407.10759}.

\bibitem[Chung et~al.(2020)Chung, Huh, Nagrani, Afouras, and Zisserman]{voxconverse}
Joon~Son Chung, Jaesung Huh, Arsha Nagrani, Triantafyllos Afouras, and Andrew Zisserman.
\newblock Spot the conversation: speaker diarisation in the wild.
\newblock 2020.

\bibitem[Cieri et~al.(2004)Cieri, Graff, Kimball, Miller, and Walker]{cieri2004fisher}
Christopher Cieri, David Graff, Owen Kimball, Dave Miller, and Kevin Walker.
\newblock Fisher english training speech part 1 transcripts.
\newblock Linguistic Data Consortium, 2004.
\newblock \url{https://catalog.ldc.upenn.edu/LDC2004T19}.
\newblock LDC2004T19.

\bibitem[D{\'e}fossez(2021)]{defossez2021hybrid}
Alexandre D{\'e}fossez.
\newblock Hybrid spectrogram and waveform source separation.
\newblock In \emph{Proceedings of the ISMIR 2021 Workshop on Music Source Separation}, 2021.

\bibitem[Dua et~al.(2025)Dua, Mittal, Gupta, and Patel]{Dua_2025}
Karan Dua, Puneet Mittal, Ranjeet Gupta, and Hitesh~Laxmichand Patel.
\newblock Speechweave: Diverse multilingual synthetic text \&amp; audio data generation pipeline for training text to speech models.
\newblock In \emph{Proceedings of the 63rd Annual Meeting of the Association for Computational Linguistics (Volume 6: Industry Track)}, page 718–737. Association for Computational Linguistics, 2025.
\newblock \doi{10.18653/v1/2025.acl-industry.51}.
\newblock \url{http://dx.doi.org/10.18653/v1/2025.acl-industry.51}.

\bibitem[Défossez et~al.(2024)Défossez, Mazaré, Orsini, Royer, Pérez, Jégou, Grave, and Zeghidour]{defossez2024moshispeechtextfoundationmodel}
Alexandre Défossez, Laurent Mazaré, Manu Orsini, Amélie Royer, Patrick Pérez, Hervé Jégou, Edouard Grave, and Neil Zeghidour.
\newblock Moshi: a speech-text foundation model for real-time dialogue, 2024.
\newblock \url{https://arxiv.org/abs/2410.00037}.

\bibitem[Fiscus(1997)]{rover}
Jonathan~G Fiscus.
\newblock A post-processing system to yield reduced word error rates: Recognizer output voting error reduction (rover).
\newblock In \emph{1997 IEEE Workshop on Automatic Speech Recognition and Understanding Proceedings}, pages 347--354. IEEE, 1997.

\bibitem[Godfrey and Holliman(1993)]{Switchboard}
John~J. Godfrey and Edward Holliman.
\newblock Switchboard-1 release 2.
\newblock Linguistic Data Consortium (LDC), 1993.
\newblock \url{https://www.ldc.upenn.edu/}.
\newblock LDC Catalog No.: LDC97S62. DOI: \url{https://doi.org/10.35111/sw3h-rw02}.

\bibitem[Goel et~al.(2025)Goel, Ghosh, Kim, Kumar, Kong, gil Lee, Yang, Duraiswami, Manocha, Valle, and Catanzaro]{goel2025audioflamingo3advancing}
Arushi Goel, Sreyan Ghosh, Jaehyeon Kim, Sonal Kumar, Zhifeng Kong, Sang gil Lee, Chao-Han~Huck Yang, Ramani Duraiswami, Dinesh Manocha, Rafael Valle, and Bryan Catanzaro.
\newblock Audio flamingo 3: Advancing audio intelligence with fully open large audio language models, 2025.
\newblock \url{https://arxiv.org/abs/2507.08128}.

\bibitem[He et~al.(2024)He, Shang, Wang, Li, Gu, Hua, Liu, Yang, Li, Shi, et~al.]{he2024emilia}
Haorui He, Zengqiang Shang, Chaoren Wang, Xuyuan Li, Yicheng Gu, Hua Hua, Liwei Liu, Chen Yang, Jiaqi Li, Peiyang Shi, et~al.
\newblock Emilia: An extensive, multilingual, and diverse speech dataset for large-scale speech generation.
\newblock In \emph{2024 IEEE Spoken Language Technology Workshop (SLT)}, pages 885--890. IEEE, 2024.

\bibitem[Hernandez et~al.(2018)Hernandez, Nguyen, Ghannay, Tomashenko, and Estève]{ted_lium}
François Hernandez, Vincent Nguyen, Sahar Ghannay, Natalia Tomashenko, and Yannick Estève.
\newblock \emph{TED-LIUM 3: Twice as Much Data and Corpus Repartition for Experiments on Speaker Adaptation}, page 198–208.
\newblock Springer International Publishing, 2018.
\newblock ISBN 9783319995793.
\newblock \doi{10.1007/978-3-319-99579-3_21}.
\newblock \url{http://dx.doi.org/10.1007/978-3-319-99579-3_21}.

\bibitem[Hurst et~al.(2024)Hurst, Lerer, Goucher, Perelman, Ramesh, Clark, Ostrow, Welihinda, Hayes, Radford, et~al.]{hurst2024gpt}
Aaron Hurst, Adam Lerer, Adam~P Goucher, Adam Perelman, Aditya Ramesh, Aidan Clark, AJ~Ostrow, Akila Welihinda, Alan Hayes, Alec Radford, et~al.
\newblock Gpt-4o system card.
\newblock \emph{arXiv preprint arXiv:2410.21276}, 2024.

\bibitem[Jung et~al.(2024{\natexlab{a}})Jung, Bae, Kim, Ryu, and Lee]{10457220}
Kyudan Jung, Seungmin Bae, Nam~Joon Kim, Hyun~Gon Ryu, and Hyuk-Jae Lee.
\newblock Improving asr performance with ocr through using word frequency difference.
\newblock In \emph{2024 International Conference on Electronics, Information, and Communication (ICEIC)}, pages 1--4, 2024{\natexlab{a}}.
\newblock \doi{10.1109/ICEIC61013.2024.10457220}.

\bibitem[Jung et~al.(2024{\natexlab{b}})Jung, Kim, Ryu, Hyeon, jun Lee, and jae Lee]{TeXBLEU}
Kyudan Jung, Nam-Joon Kim, Hyongon Ryu, Sieun Hyeon, Seung jun Lee, and Hyeok jae Lee.
\newblock Texbleu: Automatic metric for evaluate latex format, 2024{\natexlab{b}}.
\newblock \url{https://arxiv.org/abs/2409.06639}.

\bibitem[Koenecke et~al.(2024{\natexlab{a}})Koenecke, Choi, Mei, Schellmann, and Sloane]{carelessWhisper}
Allison Koenecke, Anna Seo~Gyeong Choi, Katelyn~X. Mei, Hilke Schellmann, and Mona Sloane.
\newblock Careless whisper: Speech-to-text hallucination harms.
\newblock In \emph{The 2024 ACM Conference on Fairness, Accountability, and Transparency}, FAccT ’24, page 1672–1681. ACM, June 2024{\natexlab{a}}.
\newblock \doi{10.1145/3630106.3658996}.
\newblock \url{http://dx.doi.org/10.1145/3630106.3658996}.

\bibitem[Koenecke et~al.(2024{\natexlab{b}})Koenecke, Choi, Mei, Schellmann, and Sloane]{whisperhallucination1}
Allison Koenecke, Anna Seo~Gyeong Choi, Katelyn~X Mei, Hilke Schellmann, and Mona Sloane.
\newblock Careless whisper: Speech-to-text hallucination harms.
\newblock In \emph{Proceedings of the 2024 ACM Conference on Fairness, Accountability, and Transparency}, pages 1672--1681, 2024{\natexlab{b}}.

\bibitem[Kong et~al.(2020)Kong, Cao, Iqbal, Wang, Wang, and Plumbley]{kong2020panns}
Qiuqiang Kong, Yin Cao, Turab Iqbal, Yuxuan Wang, Wenwu Wang, and Mark~D Plumbley.
\newblock Panns: Large-scale pretrained audio neural networks for audio pattern recognition.
\newblock \emph{IEEE/ACM Transactions on Audio, Speech, and Language Processing}, 28:\penalty0 2880--2894, 2020.

\bibitem[Koolagudi and Rao(2012)]{koolagudi2012emotion}
Shashidhar~G Koolagudi and K~Sreenivasa Rao.
\newblock Emotion recognition from speech: a review.
\newblock \emph{International journal of speech technology}, 15\penalty0 (2):\penalty0 99--117, 2012.

\bibitem[Lee et~al.(2025)Lee, Kim, Kim, Chung, and Cho]{lee2025dittottsdiffusiontransformersscalable}
Keon Lee, Dong~Won Kim, Jaehyeon Kim, Seungjun Chung, and Jaewoong Cho.
\newblock Ditto-tts: Diffusion transformers for scalable text-to-speech without domain-specific factors, 2025.
\newblock \url{https://arxiv.org/abs/2406.11427}.

\bibitem[Lin et~al.(2025{\natexlab{a}})Lin, Kuan, Wang, Lian, Li, and Lee]{fullduplexbench2}
Guan-Ting Lin, Shih-Yun~Shan Kuan, Qirui Wang, Jiachen Lian, Tingle Li, and Hung-yi Lee.
\newblock Full-duplex-bench v1. 5: Evaluating overlap handling for full-duplex speech models.
\newblock \emph{arXiv preprint arXiv:2507.23159}, 2025{\natexlab{a}}.

\bibitem[Lin et~al.(2025{\natexlab{b}})Lin, Lian, Li, Wang, Anumanchipalli, Liu, and Lee]{fullduplexbench1}
Guan-Ting Lin, Jiachen Lian, Tingle Li, Qirui Wang, Gopala Anumanchipalli, Alexander~H Liu, and Hung-yi Lee.
\newblock Full-duplex-bench: A benchmark to evaluate full-duplex spoken dialogue models on turn-taking capabilities.
\newblock \emph{arXiv preprint arXiv:2503.04721}, 2025{\natexlab{b}}.

\bibitem[Mansoor et~al.(2025)Mansoor, Abdullah, Adil, Jamil, Hameed, and Soleimani]{whisperhallucination3}
Harras Mansoor, Umer Abdullah, Shahryar Adil, Akhtar Jamil, Alaa~Ali Hameed, and Faezeh Soleimani.
\newblock Mitigating hallucinations in speech recognition systems for noisy data.
\newblock In \emph{2025 IEEE 4th International Conference on Computing and Machine Intelligence (ICMI)}, pages 1--5. IEEE, 2025.

\bibitem[McFee(2025)]{mcfee2025librosa}
Brian McFee.
\newblock librosa/librosa: 0.11.0, March 2025.
\newblock \url{https://doi.org/10.5281/zenodo.15006942}.
\newblock \textnormal{Version 0.11.0}.

\bibitem[Panayotov et~al.(2015)Panayotov, Chen, Povey, and Khudanpur]{librispeech}
Vassil Panayotov, Guoguo Chen, Daniel Povey, and Sanjeev Khudanpur.
\newblock Librispeech: An asr corpus based on public domain audio books.
\newblock In \emph{2015 IEEE International Conference on Acoustics, Speech and Signal Processing (ICASSP)}, pages 5206--5210, 2015.
\newblock \doi{10.1109/ICASSP.2015.7178964}.

\bibitem[Park et~al.()Park, Medennikov, Dhawan, Wang, Huang, Koluguri, Puvvada, Balam, and Ginsburg]{parksortformer}
Taejin Park, Ivan Medennikov, Kunal Dhawan, Weiqing Wang, He~Huang, Nithin~Rao Koluguri, Krishna~C Puvvada, Jagadeesh Balam, and Boris Ginsburg.
\newblock Sortformer: A novel approach for permutation-resolved speaker supervision in speech-to-text systems.
\newblock In \emph{Forty-second International Conference on Machine Learning}.

\bibitem[Penedo et~al.(2024)Penedo, Kydl{\'\i}{\v{c}}ek, Lozhkov, Mitchell, Raffel, Von~Werra, Wolf, et~al.]{fineweb}
Guilherme Penedo, Hynek Kydl{\'\i}{\v{c}}ek, Anton Lozhkov, Margaret Mitchell, Colin~A Raffel, Leandro Von~Werra, Thomas Wolf, et~al.
\newblock The fineweb datasets: Decanting the web for the finest text data at scale.
\newblock \emph{Advances in Neural Information Processing Systems}, 37:\penalty0 30811--30849, 2024.

\bibitem[Plaquet and Bredin(2023)]{pyannote2}
Alexis Plaquet and Hervé Bredin.
\newblock {Powerset multi-class cross entropy loss for neural speaker diarization}.
\newblock In \emph{Proc. INTERSPEECH 2023}, 2023.

\bibitem[Radford et~al.(2022{\natexlab{a}})Radford, Kim, Xu, Brockman, McLeavey, and Sutskever]{radford2022robustspeechrecognitionlargescale}
Alec Radford, Jong~Wook Kim, Tao Xu, Greg Brockman, Christine McLeavey, and Ilya Sutskever.
\newblock Robust speech recognition via large-scale weak supervision, 2022{\natexlab{a}}.
\newblock \url{https://arxiv.org/abs/2212.04356}.

\bibitem[Radford et~al.(2022{\natexlab{b}})Radford, Kim, Xu, Brockman, McLeavey, and Sutskever]{radford2022whisper}
Alec Radford, Jong~Wook Kim, Tao Xu, Greg Brockman, Christine McLeavey, and Ilya Sutskever.
\newblock Robust speech recognition via large-scale weak supervision, 2022{\natexlab{b}}.
\newblock \url{https://arxiv.org/abs/2212.04356}.

\bibitem[Rouard et~al.(2023)Rouard, Massa, and D{\'e}fossez]{rouard2022hybrid}
Simon Rouard, Francisco Massa, and Alexandre D{\'e}fossez.
\newblock Hybrid transformers for music source separation.
\newblock In \emph{ICASSP 23}, 2023.

\bibitem[Roy et~al.(2026)Roy, Raiman, Lee, Ene, Kirby, Kim, Kim, and Catanzaro]{roy2026personaplex}
Rajarshi Roy, Jonathan Raiman, Sang-gil Lee, Teodor-Dumitru Ene, Robert Kirby, Sungwon Kim, Jaehyeon Kim, and Bryan Catanzaro.
\newblock Personaplex: Voice and role control for full duplex conversational speech models.
\newblock 2026.

\bibitem[Rubenstein et~al.(2023)Rubenstein, Asawaroengchai, Nguyen, Bapna, Borsos, de~Chaumont~Quitry, Chen, Badawy, Han, Kharitonov, Muckenhirn, Padfield, Qin, Rozenberg, Sainath, Schalkwyk, Sharifi, Ramanovich, Tagliasacchi, Tudor, Velimirović, Vincent, Yu, Wang, Zayats, Zeghidour, Zhang, Zhang, Zilka, and Frank]{rubenstein2023audiopalmlargelanguagemodel}
Paul~K. Rubenstein, Chulayuth Asawaroengchai, Duc~Dung Nguyen, Ankur Bapna, Zalán Borsos, Félix de~Chaumont~Quitry, Peter Chen, Dalia~El Badawy, Wei Han, Eugene Kharitonov, Hannah Muckenhirn, Dirk Padfield, James Qin, Danny Rozenberg, Tara Sainath, Johan Schalkwyk, Matt Sharifi, Michelle~Tadmor Ramanovich, Marco Tagliasacchi, Alexandru Tudor, Mihajlo Velimirović, Damien Vincent, Jiahui Yu, Yongqiang Wang, Vicky Zayats, Neil Zeghidour, Yu~Zhang, Zhishuai Zhang, Lukas Zilka, and Christian Frank.
\newblock Audiopalm: A large language model that can speak and listen, 2023.
\newblock \url{https://arxiv.org/abs/2306.12925}.

\bibitem[Saeki et~al.(2022)Saeki, Xin, Nakata, Koriyama, Takamichi, and Saruwatari]{saeki2022utmos}
Takaaki Saeki, Detai Xin, Wataru Nakata, Tomoki Koriyama, Shinnosuke Takamichi, and Hiroshi Saruwatari.
\newblock Utmos: Utokyo-sarulab system for voicemos challenge 2022.
\newblock \emph{arXiv preprint arXiv:2204.02152}, 2022.

\bibitem[Sekoyan et~al.(2025)Sekoyan, Koluguri, Tadevosyan, Zelasko, Bartley, Karpov, Balam, and Ginsburg]{sekoyan2025canary1bv2parakeettdt06bv3efficient}
Monica Sekoyan, Nithin~Rao Koluguri, Nune Tadevosyan, Piotr Zelasko, Travis Bartley, Nikolay Karpov, Jagadeesh Balam, and Boris Ginsburg.
\newblock Canary-1b-v2 \& parakeet-tdt-0.6b-v3: Efficient and high-performance models for multilingual asr and ast, 2025.
\newblock \url{https://arxiv.org/abs/2509.14128}.

\bibitem[Shi et~al.(2025)Shi, Tjandra, Hoffman, Wang, Wu, Gao, Richter, Le, Vyas, Chen, Feichtenhofer, Dollár, Hsu, and Lee]{shi2025samaudiosegmentaudio}
Bowen Shi, Andros Tjandra, John Hoffman, Helin Wang, Yi-Chiao Wu, Luya Gao, Julius Richter, Matt Le, Apoorv Vyas, Sanyuan Chen, Christoph Feichtenhofer, Piotr Dollár, Wei-Ning Hsu, and Ann Lee.
\newblock Sam audio: Segment anything in audio, 2025.
\newblock \url{https://arxiv.org/abs/2512.18099}.

\bibitem[Shin et~al.(2024)Shin, Lee, Kim, and Park]{shin2024separate}
Ui-Hyeop Shin, Sangyoun Lee, Taehan Kim, and Hyung-Min Park.
\newblock Separate and reconstruct: Asymmetric encoder-decoder for speech separation.
\newblock \emph{Advances in Neural Information Processing Systems}, 37:\penalty0 52215--52240, 2024.

\bibitem[Soldaini et~al.(2024)Soldaini, Kinney, Bhagia, Schwenk, Atkinson, Authur, Bogin, Chandu, Dumas, Elazar, et~al.]{dolma}
Luca Soldaini, Rodney Kinney, Akshita Bhagia, Dustin Schwenk, David Atkinson, Russell Authur, Ben Bogin, Khyathi Chandu, Jennifer Dumas, Yanai Elazar, et~al.
\newblock Dolma: An open corpus of three trillion tokens for language model pretraining research.
\newblock In \emph{Proceedings of the 62nd annual meeting of the association for computational linguistics (volume 1: long papers)}, pages 15725--15788, 2024.

\bibitem[Team(2026{\natexlab{a}})]{navercloudhyperclovaxteam2026hyperclovax32bthink}
NAVER Cloud HyperCLOVA~X Team.
\newblock Hyperclova x 32b think, 2026{\natexlab{a}}.
\newblock \url{https://arxiv.org/abs/2601.03286}.

\bibitem[Team(2026{\natexlab{b}})]{navercloudhyperclovaxteam2026hyperclovax8bomni}
NAVER Cloud HyperCLOVA~X Team.
\newblock Hyperclova x 8b omni, 2026{\natexlab{b}}.
\newblock \url{https://arxiv.org/abs/2601.01792}.

\bibitem[Team(2024)]{SileroVAD}
Silero Team.
\newblock Silero vad: pre-trained enterprise-grade voice activity detector (vad), number detector and language classifier.
\newblock \url{https://github.com/snakers4/silero-vad}, 2024.

\bibitem[Tursunov et~al.(2019)Tursunov, Kwon, and Pang]{tursunov2019discriminating}
Anvarjon Tursunov, Soonil Kwon, and Hee-Suk Pang.
\newblock Discriminating emotions in the valence dimension from speech using timbre features.
\newblock \emph{Applied Sciences}, 9\penalty0 (12):\penalty0 2470, 2019.

\bibitem[Udandarao et~al.(2025)Udandarao, Lu, Chang, Wang, Yao, Jose, Faghri, Gardner, and Chiu]{udandarao2025datacentriclessonsimprovespeechlanguage}
Vishaal Udandarao, Zhiyun Lu, Xuankai Chang, Yongqiang Wang, Violet~Z. Yao, Albin~Madapally Jose, Fartash Faghri, Josh Gardner, and Chung-Cheng Chiu.
\newblock Data-centric lessons to improve speech-language pretraining, 2025.
\newblock \url{https://arxiv.org/abs/2510.20860}.

\bibitem[Wang et~al.(2025{\natexlab{a}})Wang, Hai, Chong, Thakkar, Feng, Yang, Lee, Thebaud, Velazquez, Villalba, et~al.]{wang2025capspeech}
Helin Wang, Jiarui Hai, Dading Chong, Karan Thakkar, Tiantian Feng, Dongchao Yang, Junhyeok Lee, Thomas Thebaud, Laureano~Moro Velazquez, Jesus Villalba, et~al.
\newblock Capspeech: Enabling downstream applications in style-captioned text-to-speech.
\newblock \emph{arXiv preprint arXiv:2506.02863}, 2025{\natexlab{a}}.

\bibitem[Wang et~al.(2024{\natexlab{a}})Wang, Chen, Khare, Raju, Dheram, He, Wu, Stolcke, and Ravichandran]{wang2024turn}
Jinhan Wang, Long Chen, Aparna Khare, Anirudh Raju, Pranav Dheram, Di~He, Minhua Wu, Andreas Stolcke, and Venkatesh Ravichandran.
\newblock Turn-taking and backchannel prediction with acoustic and large language model fusion.
\newblock In \emph{ICASSP 2024-2024 IEEE International Conference on Acoustics, Speech and Signal Processing (ICASSP)}, pages 12121--12125. IEEE, 2024{\natexlab{a}}.

\bibitem[Wang et~al.(2024{\natexlab{b}})Wang, Lu, Tang, Yan, Xia, and Xiong]{wang2024fullduplexspeechdialoguescheme}
Peng Wang, Songshuo Lu, Yaohua Tang, Sijie Yan, Wei Xia, and Yuanjun Xiong.
\newblock A full-duplex speech dialogue scheme based on large language models, 2024{\natexlab{b}}.
\newblock \url{https://arxiv.org/abs/2405.19487}.

\bibitem[Wang et~al.(2025{\natexlab{b}})Wang, Chen, Lin, Raj, Kimball, Cabral, and Hester]{wang2025companioncastmultiagentconversationalai}
Yiyang Wang, Chen Chen, Tica Lin, Vishnu Raj, Josh Kimball, Alex Cabral, and Josiah Hester.
\newblock Companioncast: A multi-agent conversational ai framework with spatial audio for social co-viewing experiences, 2025{\natexlab{b}}.
\newblock \url{https://arxiv.org/abs/2512.10918}.

\bibitem[Weber et~al.(2024)Weber, Fu, Anthony, Oren, Adams, Alexandrov, Lyu, Nguyen, Yao, Adams, et~al.]{weber2024redpajama}
Maurice Weber, Dan Fu, Quentin Anthony, Yonatan Oren, Shane Adams, Anton Alexandrov, Xiaozhong Lyu, Huu Nguyen, Xiaozhe Yao, Virginia Adams, et~al.
\newblock Redpajama: an open dataset for training large language models.
\newblock \emph{Advances in neural information processing systems}, 37:\penalty0 116462--116492, 2024.

\bibitem[Xu et~al.(2025{\natexlab{a}})Xu, Guo, He, Hu, He, Bai, Chen, Wang, Fan, Dang, Zhang, Wang, Chu, and Lin]{xu2025qwen25omnitechnicalreport}
Jin Xu, Zhifang Guo, Jinzheng He, Hangrui Hu, Ting He, Shuai Bai, Keqin Chen, Jialin Wang, Yang Fan, Kai Dang, Bin Zhang, Xiong Wang, Yunfei Chu, and Junyang Lin.
\newblock Qwen2.5-omni technical report, 2025{\natexlab{a}}.
\newblock \url{https://arxiv.org/abs/2503.20215}.

\bibitem[Xu et~al.(2025{\natexlab{b}})Xu, Guo, Hu, Chu, Wang, He, Wang, Shi, He, Zhu, Lv, Wang, Guo, Wang, Ma, Zhang, Zhang, Hao, Guo, Yang, Zhang, Ma, Wei, Bai, Chen, Liu, Wang, Yang, Liu, Ren, Zheng, Men, Zhou, Yu, Yang, Yu, Zhou, and Lin]{xu2025qwen3omnitechnicalreport}
Jin Xu, Zhifang Guo, Hangrui Hu, Yunfei Chu, Xiong Wang, Jinzheng He, Yuxuan Wang, Xian Shi, Ting He, Xinfa Zhu, Yuanjun Lv, Yongqi Wang, Dake Guo, He~Wang, Linhan Ma, Pei Zhang, Xinyu Zhang, Hongkun Hao, Zishan Guo, Baosong Yang, Bin Zhang, Ziyang Ma, Xipin Wei, Shuai Bai, Keqin Chen, Xuejing Liu, Peng Wang, Mingkun Yang, Dayiheng Liu, Xingzhang Ren, Bo~Zheng, Rui Men, Fan Zhou, Bowen Yu, Jianxin Yang, Le~Yu, Jingren Zhou, and Junyang Lin.
\newblock Qwen3-omni technical report, 2025{\natexlab{b}}.
\newblock \url{https://arxiv.org/abs/2509.17765}.

\bibitem[Yan et~al.(2025)Yan, Jin, Huang, Yu, Peng, Zhan, Gao, Peng, Chen, Zhou, Ren, Yang, Yang, Xu, Zhao, Xiong, Lin, Wang, Yuan, Wu, Lyu, He, Qiu, Fang, and Huang]{yan2025minguniaudiospeechllmjoint}
Canxiang Yan, Chunxiang Jin, Dawei Huang, Haibing Yu, Han Peng, Hui Zhan, Jie Gao, Jing Peng, Jingdong Chen, Jun Zhou, Kaimeng Ren, Ming Yang, Mingxue Yang, Qiang Xu, Qin Zhao, Ruijie Xiong, Shaoxiong Lin, Xuezhi Wang, Yi~Yuan, Yifei Wu, Yongjie Lyu, Zhengyu He, Zhihao Qiu, Zhiqiang Fang, and Ziyuan Huang.
\newblock Ming-uniaudio: Speech llm for joint understanding, generation and editing with unified representation, 2025.
\newblock \url{https://arxiv.org/abs/2511.05516}.

\bibitem[Yang et~al.(2025)Yang, Song, Zhuo, Cui, Li, Yang, Du, Ma, Liu, Wang, Li, Fan, Yu, Zhang, Chen, and Chen]{yang2025gigaspeech2evolvinglargescale}
Yifan Yang, Zheshu Song, Jianheng Zhuo, Mingyu Cui, Jinpeng Li, Bo~Yang, Yexing Du, Ziyang Ma, Xunying Liu, Ziyuan Wang, Ke~Li, Shuai Fan, Kai Yu, Wei-Qiang Zhang, Guoguo Chen, and Xie Chen.
\newblock Gigaspeech 2: An evolving, large-scale and multi-domain asr corpus for low-resource languages with automated crawling, transcription and refinement, 2025.
\newblock \url{https://arxiv.org/abs/2406.11546}.

\bibitem[Ye et~al.(2025)Ye, Yang, Goel, Huang, Zhu, Su, Lin, Cheng, Wan, Tian, et~al.]{ye2025omnivinci}
Hanrong Ye, Chao-Han~Huck Yang, Arushi Goel, Wei Huang, Ligeng Zhu, Yuanhang Su, Sean Lin, An-Chieh Cheng, Zhen Wan, Jinchuan Tian, et~al.
\newblock Omnivinci: Enhancing architecture and data for omni-modal understanding llm.
\newblock \emph{arXiv preprint arXiv:2510.15870}, 2025.

\bibitem[Zhang et~al.(2022)Zhang, Lv, Guo, Shao, Yang, Xie, Xu, Bu, Chen, Zeng, Wu, and Peng]{zhang2022wenetspeech10000hoursmultidomain}
Binbin Zhang, Hang Lv, Pengcheng Guo, Qijie Shao, Chao Yang, Lei Xie, Xin Xu, Hui Bu, Xiaoyu Chen, Chenchen Zeng, Di~Wu, and Zhendong Peng.
\newblock Wenetspeech: A 10000+ hours multi-domain mandarin corpus for speech recognition, 2022.
\newblock \url{https://arxiv.org/abs/2110.03370}.

\end{thebibliography}

\clearpage
\newpage

\appendix
\section*{Appendix}
We provide detailed supplementary materials organized as follows:

\begin{itemize} \item \textbf{Appendix}~\ref{related_works} reviews related work on full-duplex models, large-scale speech datasets, and speech preprocessing techniques. \item \textbf{Appendix}~\ref{app:detail_overlap} presents additional experimental results regarding overlap separation. 
\item \textbf{Appendix}~\ref{app:overlap_cases} illustrates specific cases for handling backchanneling and overlapping speech. 
\item \textbf{Appendix}~\ref{app:detail_of_finetuning} details supplementary results from the fine-tuning experiments. 
\item \textbf{Appendix}~\ref{context_captioning} describes the techniques employed for audio captioning. 
\item \textbf{Appendix}~\ref{example} provides examples of data processed by \textsc{Sommelier}.
\end{itemize}
\hrulefill
\vspace{1em}

\section{Related Works}
\label{related_works}
In this section, we provide a comprehensive overview of the research landscape relevant to our work. 
We begin by tracing the evolution of Speech Language Models from cascaded pipelines to end-to-end architectures, with a particular focus on the recent shift toward full-duplex interaction (\S\ref{sec:slm}). 
We then critically examine the current landscape of large-scale speech datasets, identifying their structural limitations for modeling naturalistic conversational dynamics (\S\ref{sec:datasets}). 
Finally, we survey the automated pipelines used for speech data curation, discussing key technical challenges such as speaker diarization errors and transcription hallucinations that our work aims to address (\S\ref{sec:pipelines}).

\subsection{Speech Language Models and Full-Duplex Interaction}
\label{sec:slm}
The landscape of spoken language understanding has undergone a fundamental transformation, moving away from cascaded systems towards end-to-end modeling.
Traditional conversational agents relied on a cascade pipeline of Automatic Speech Recognition (ASR), Large Language Models (LLM), and Text-to-Speech (TTS)~\citep{lee2025dittottsdiffusiontransformersscalable, 10457220}.
While effective for distinct tasks, this modular approach inevitably suffers from error propagation and the loss of paralinguistic features such as emotion, prosody, and intonationp.
To address these limitations, End-to-End Speech Language Models (SLMs), such as AudioLM~\citep{borsos2023audiolmlanguagemodelingapproach} and AudioPaLM~\citep{rubenstein2023audiopalmlargelanguagemodel}, were introduced to process acoustic tokens directly, preserving the rich nuances of speech.
Building on this foundation, recent multimodal models like Qwen-Audio~\cite{chu2023qwenaudioadvancinguniversalaudio, chu2024qwen2audiotechnicalreport}, Qwen-Omni~\cite{xu2025qwen25omnitechnicalreport, xu2025qwen3omnitechnicalreport}, HyerCLOVA-X-8B-Omni~\citep{navercloudhyperclovaxteam2026hyperclovax8bomni, navercloudhyperclovaxteam2026hyperclovax32bthink} and Audio Flamingo~\cite{chu2024qwen2audiotechnicalreport, goel2025audioflamingo3advancing} have demonstrated exceptional capabilities in understanding and reasoning across both text and audio modalities, effectively bridging the gap between sound processing and language comprehension.

Despite these advancements, a critical frontier remains in achieving natural, real-time human-computer interaction.
While earlier models primarily operated on a turn-based (half-duplex) mechanism, recent developments like Moshi~\cite{defossez2024moshispeechtextfoundationmodel} and GPT-4o~\cite{hurst2024gpt} aim to realize full-duplex communication, where listening and speaking occur simultaneously.
This shift necessitates models to master complex conversational dynamics, including the ability to handle overlapping speech, process back-channeling (e.g., rapid affirmations), and predict turn-taking opportunities seamlessly.
However, training such full-duplex systems presents a significant challenge: it requires high-quality, multi-stream data that captures these intricate acoustic intersections.
The scarcity of such datasets in the current research landscape creates a bottleneck, limiting the ability of current models to generalize to the chaotic and fluid nature of real-world dialogue.

\subsection{Large-Scale Speech Datasets}
\label{sec:datasets}
Existing speech datasets, despite their increasing volume, remain suboptimal for training full-duplex models that require rich interactional dynamics.
Traditional ASR benchmarks like LibriSpeech~\citep{librispeech} and GigaSpeech~\citep{gigaspeech} are dominated by scripted read speech or solitary monologues, failing to capture the dynamic and interactive spontaneity of human dialogue.
While conversational datasets such as Fisher~\citep{cieri2004fisher} and Switchboard~\citep{Switchboard} offer multi-speaker interactions, they are severely limited by their archaic telephony quality (8kHz), narrow bandwidth, and relatively small scale (typically a few thousand hours), which is insufficient for modern large-scale pre-training~\cite{radford2022whisper,xu2025qwen3omnitechnicalreport}.
While recent web-scale initiatives like WenetSpeech~\citep{zhang2022wenetspeech10000hoursmultidomain} and Emilia~\cite{he2024emilia} have successfully aggregated massive datasets, their pipelines are heavily optimized for single-stream speech, thereby neglecting the concurrent dynamics required for full-duplex interaction.
Crucially, their pre-processing pipelines  treat overlapping speech as noise to be excised or ignored rather than a feature to be modeled.
This structural limitation results in data that lacks the distinct multi-stream separation and essential acoustic collisions required for learning true full-duplex interaction.

\subsection{Automated Speech Data Processing Pipelines}
\label{sec:pipelines}

While open-source data processing pipelines have become the bedrock of Large Language Model (LLM) research, exemplified by transparent frameworks like Dolma~\citep{dolma}, RedPajama~\citep{weber2024redpajama}, and FineWeb~\citep{fineweb}, the domain of speech processing remains significantly opaque.
Although model weights for Speech Language Models (SLMs) are frequently released, the intricate ``data recipes'' required to curate high-quality pre-training corpora remain proprietary ``black boxes,'' impeding the community's ability to reproduce results or improve upon existing strategies.

This lack of standardized, open pipelines is particularly critical when addressing the technical demands of full-duplex communication.
Current methodologies rely heavily on tools designed for single-stream processing~\citep{Dua_2025, yang2025gigaspeech2evolvinglargescale}, which are ill-suited for capturing the concurrent dynamics of human dialogue.
For instance, while speaker diarization is a prerequisite for multi-turn modeling, standard tools like Pyannote~\citep{pyannote1, pyannote2} often struggle in the complex acoustic environments of in-the-wild web videos.
Crucially, these tools frequently misinterpret the \textit{overlaps} and rapid \textit{turn-taking}, essential features of full-duplex interaction, as segmentation errors or noise, thereby degrading the structural integrity of the conversational data.

Furthermore, the reliance on ASR models for transcription introduces the risk of hallucinations. Models like Whisper~\citep{radford2022whisper}, though powerful, are prone to generating repetitive loops or nonsensical text during silence or non-speech intervals, a critical instability highlighted in recent studies such as \textit{Careless Whisper}~\citep{carelessWhisper}.
Existing pipelines lack the robustness to filter these hallucinations or handle the multi-stream nature of duplex speech, underscoring the urgent need for a transparent, hallucination-aware processing framework tailored for conversational AI.

\section{Detail of Overlapping disentangle experiments}
\label{app:detail_overlap}

\begin{figure*}[t]
    \centering
    \includegraphics[width=1\linewidth]{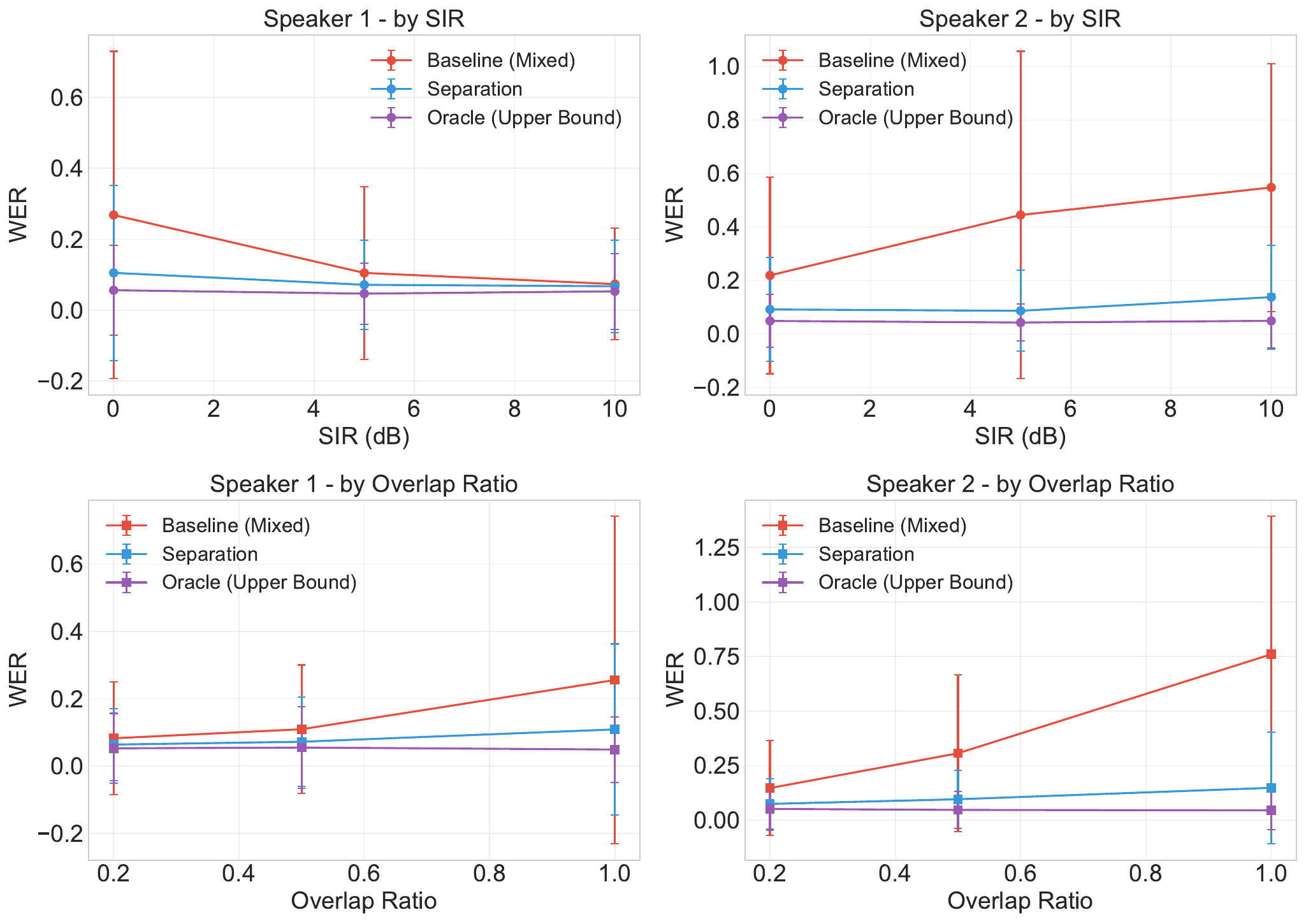}
    \caption{WER comparison by method, SIR, and overlap ratio for both speakers. Top: WER as a function of SIR (dB). Bottom: WER as a function of overlap ratio. Methods include Baseline (mixed), Separation, and Oracle. Error bars represent standard deviation.}
    \label{fig:graph_overlap}
\end{figure*}

This section presents detailed experimental results on the efficacy of our overlap separation module.
We evaluated performance across varying Signal-to-Interference Ratios (SIR) and Overlap Ratios using four key metrics: Word Error Rate (WER), Scale-Invariant Signal-to-Distortion Ratio (SI-SDR), Short-Time Objective Intelligibility (STOI), and UTMOS.
The results, summarized in Tables~\ref{tab:wer} through \ref{tab:utmos}, demonstrate that applying the separation module (`Sep') consistently improves signal quality compared to the baseline (`Base').
Notably, the performance gain is significantly larger for Speaker 2 (the interfering or secondary speaker) than for Speaker 1 (the primary speaker), particularly in challenging conditions with high overlap ratios as shown in Figure~\ref{fig:graph_overlap}.

\subsection{Analysis of Results}

\paragraph{Asymmetric Gains (Spk1 vs. Spk2)}
Across all metrics, the gap between the `Base' and `Sep' conditions is most dramatic for Speaker 2. For instance, in the 0,dB SIR and 1.0 overlap condition, the WER for Speaker 2 improves drastically from 0.444 to 0.138 (Table~\ref{tab:wer}), whereas Speaker 1 sees a relatively smaller, though still significant, improvement. This suggests that our module is particularly effective at recovering the subordinate or quieter speaker in a mixture, which is crucial for full-duplex conversational AI where both parties must be heard clearly.
\paragraph{Resilience to High Overlap}
The benefits of separation become more pronounced as the overlap ratio increases. In the worst-case scenario (1.0 overlap), the baseline UTMOS scores (Table~\ref{tab:utmos}) drop severely (e.g., $\approx$1.7), but the separation module restores quality to near-natural levels ($\approx$3.0). Similarly, STOI scores (Table~\ref{tab:stoi}) remain high ($>$0.9) even under full overlap when separation is applied, confirming that intelligibility is preserved.

% ============================================
% Table 1: WER Results (↓ lower is better)
% ============================================
\begin{table}[t]
\centering

\resizebox{0.6\textwidth}{!}{%
\begin{tabular}{cc|ccc|ccc}
\toprule
\multirow{2}{*}{\opthead{SIR}} & \multirow{2}{*}{\opthead{Overlap}} & \multicolumn{3}{c|}{\opthead{Speaker 1}} & \multicolumn{3}{c}{\opthead{Speaker 2}} \\
 & & Base & Sep & Oracle & Base & Sep & Oracle \\
\midrule
\multirow{3}{*}{0 dB} 
 & 0.2 & 0.109 & \textbf{0.048} & 0.034 & 0.100 & \textbf{0.074} & 0.061 \\
 & 0.5 & 0.162 & \textbf{0.094} & 0.078 & 0.115 & \textbf{0.065} & 0.039 \\
 & 1.0 & 0.535 & \textbf{0.175} & 0.058 & 0.444 & \textbf{0.138} & 0.048 \\
\midrule
\multirow{3}{*}{5 dB} 
 & 0.2 & \textbf{0.080} & 0.088 & 0.066 & 0.146 & \textbf{0.065} & 0.040 \\
 & 0.5 & 0.099 & \textbf{0.059} & 0.036 & 0.277 & \textbf{0.084} & 0.049 \\
 & 1.0 & 0.136 & \textbf{0.069} & 0.039 & 0.913 & \textbf{0.113} & 0.042 \\
\midrule
\multirow{3}{*}{10 dB} 
 & 0.2 & 0.059 & \textbf{0.056} & 0.058 & 0.194 & \textbf{0.084} & 0.053 \\
 & 0.5 & 0.067 & \textbf{0.064} & 0.051 & 0.527 & \textbf{0.138} & 0.052 \\
 & 1.0 & 0.096 & \textbf{0.083} & 0.051 & 0.923 & \textbf{0.193} & 0.044 \\
\bottomrule
\end{tabular}}
\caption{Word Error Rate (WER) comparison across different conditions. Lower is better.}
\label{tab:wer}
\end{table}

% ============================================
% Table 3: STOI Results (↑ higher is better)
% ============================================
\begin{table}[t]
\centering
\resizebox{0.6\textwidth}{!}{%
\begin{tabular}{cc|ccc|ccc}
\toprule
\multirow{2}{*}{\opthead{SIR}} & \multirow{2}{*}{\opthead{Overlap}} & \multicolumn{3}{c|}{\opthead{Speaker 1}} & \multicolumn{3}{c}{\opthead{Speaker 2}} \\
 & & Base & Sep & Oracle & Base & Sep & Oracle \\
\midrule
\multirow{3}{*}{0 dB} 
 & 0.2 & 0.959 & \textbf{0.980} & 1.000 & 0.964 & \textbf{0.985} & 1.000 \\
 & 0.5 & 0.889 & \textbf{0.969} & 1.000 & 0.887 & \textbf{0.968} & 1.000 \\
 & 1.0 & 0.785 & \textbf{0.918} & 1.000 & 0.771 & \textbf{0.908} & 1.000 \\
\midrule
\multirow{3}{*}{5 dB} 
 & 0.2 & 0.971 & \textbf{0.980} & 1.000 & 0.951 & \textbf{0.976} & 1.000 \\
 & 0.5 & 0.929 & \textbf{0.980} & 1.000 & 0.844 & \textbf{0.962} & 1.000 \\
 & 1.0 & 0.847 & \textbf{0.955} & 1.000 & 0.676 & \textbf{0.917} & 1.000 \\
\midrule
\multirow{3}{*}{10 dB} 
 & 0.2 & 0.985 & \textbf{0.988} & 1.000 & 0.938 & \textbf{0.971} & 1.000 \\
 & 0.5 & 0.956 & \textbf{0.981} & 1.000 & 0.798 & \textbf{0.931} & 1.000 \\
 & 1.0 & 0.901 & \textbf{0.954} & 1.000 & 0.608 & \textbf{0.883} & 1.000 \\
\bottomrule
\end{tabular}}
\caption{Short-Time Objective Intelligibility (STOI) scores. Higher is better.}
\label{tab:stoi}
\end{table}

% ============================================
% Table 4: UTMOS Results (↑ higher is better)
% ============================================
\begin{table}[t]
\centering

\resizebox{0.6\textwidth}{!}{%
\begin{tabular}{cc|ccc|ccc}
\toprule
\multirow{2}{*}{\opthead{SIR}} & \multirow{2}{*}{\opthead{Overlap}} & \multicolumn{3}{c|}{\opthead{Speaker 1}} & \multicolumn{3}{c}{\opthead{Speaker 2}} \\
 & & Base & Sep & Oracle & Base & Sep & Oracle \\
\midrule
\multirow{3}{*}{0 dB} 
 & 0.2 & 3.08 & \textbf{3.56} & 3.94 & 2.99 & \textbf{3.50} & 3.82 \\
 & 0.5 & 2.28 & \textbf{3.29} & 3.85 & 2.25 & \textbf{3.35} & 3.90 \\
 & 1.0 & 1.73 & \textbf{3.05} & 3.81 & 1.67 & \textbf{2.99} & 3.86 \\
\midrule
\multirow{3}{*}{5 dB} 
 & 0.2 & 3.12 & \textbf{3.47} & 3.83 & 3.00 & \textbf{3.48} & 3.90 \\
 & 0.5 & 2.46 & \textbf{3.51} & 3.90 & 2.22 & \textbf{3.28} & 3.92 \\
 & 1.0 & 1.86 & \textbf{3.29} & 3.94 & 1.72 & \textbf{2.94} & 3.87 \\
\midrule
\multirow{3}{*}{10 dB} 
 & 0.2 & 3.39 & \textbf{3.69} & 3.95 & 3.12 & \textbf{3.51} & 4.02 \\
 & 0.5 & 2.82 & \textbf{3.50} & 3.91 & 2.35 & \textbf{2.92} & 3.82 \\
 & 1.0 & 2.22 & \textbf{3.29} & 3.95 & 2.12 & \textbf{2.73} & 3.90 \\
\bottomrule
\end{tabular}}
\caption{UTMOS (MOS prediction) scores. Higher is better.}
\label{tab:utmos}
\end{table}

the Baseline method exhibits competitive SI-SDR performance at $\rho \in \{0.2, 0.5\}$. However, this comparison is inherently unfair: the Baseline extracts each speaker's time segment directly from the mixed signal, meaning that at 20\% and 50\% overlap ratios, the majority of each segment contains clean, non-overlapping speech. For instance, at $\rho=0.2$, the actual overlap constitutes only 15.2\% for S1 and 14.7\% for S2, leaving approximately 7 seconds of clean audio per speaker. Consequently, ASR systems can effectively recognize the clean portions, resulting in artificially low WER for the Baseline.

\section{Overlap cases}
\label{app:overlap_cases}
In this section, we present two representative overlap cases. The first is \textit{backchanneling}, where a speaker produces a short utterance while the other is speaking; in this case, one segment is fully contained within another segment. The second is \textit{overlap}, where two segments partially overlap but neither segment fully contains the other.
\begin{figure}[H]
    \centering
    \includegraphics[width=0.4\linewidth]{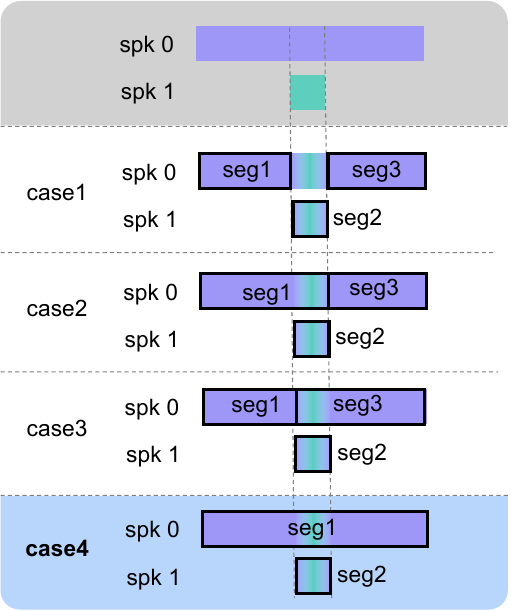}
    \caption{Four ways to handle backchanneling in overlapping speech.}
    \label{fig:overlap-backchanneling}
\end{figure}
\begin{figure}[H]
    \centering
    \includegraphics[width=0.4\linewidth, trim={0 0 0 0}, clip]{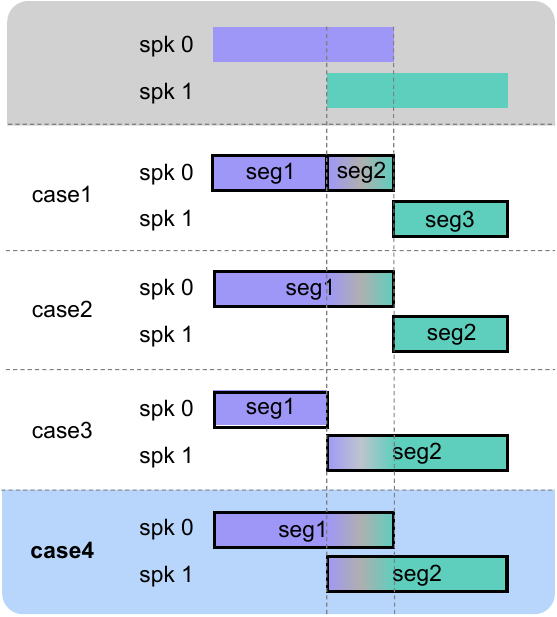}
    \caption{Four distinct types of separable cases in overlapping speech.}
    \label{fig:overlap}
\end{figure}

\section{Detail of Finetuning Experiment}
\label{app:detail_of_finetuning}

The fine-tuning hyperparameters for Moshiko are listed in Table~\ref{training_config}. We significantly benefited from the implementation provided at \url{https://github.com/kyutai-labs/moshi-finetune}.
\begin{table}[h]
    \centering
    \begin{tabular}{lc}
        \toprule
        \opthead{Hyperparameter} & \opthead{Value} \\
        \midrule
        Total Data Duration & $\approx$ 83 hours \\
        Training Steps & 2,000 \\
        Hardware & $8 \times \text{A100}$ \\
        Rank & 128 \\
        Batch Size & 16 \\
        Learning Rate & $2\mathrm{e}{-6}$ \\
        Weight Decay & 0.1 \\
        \bottomrule
    \end{tabular}
    \caption{Training hyperparameters and settings.}
    \label{training_config}
\end{table}

\subsection{Dataset Statistics}
\label{dataset statistics}
\begin{figure*}
    \centering
    \includegraphics[width=0.65\linewidth]{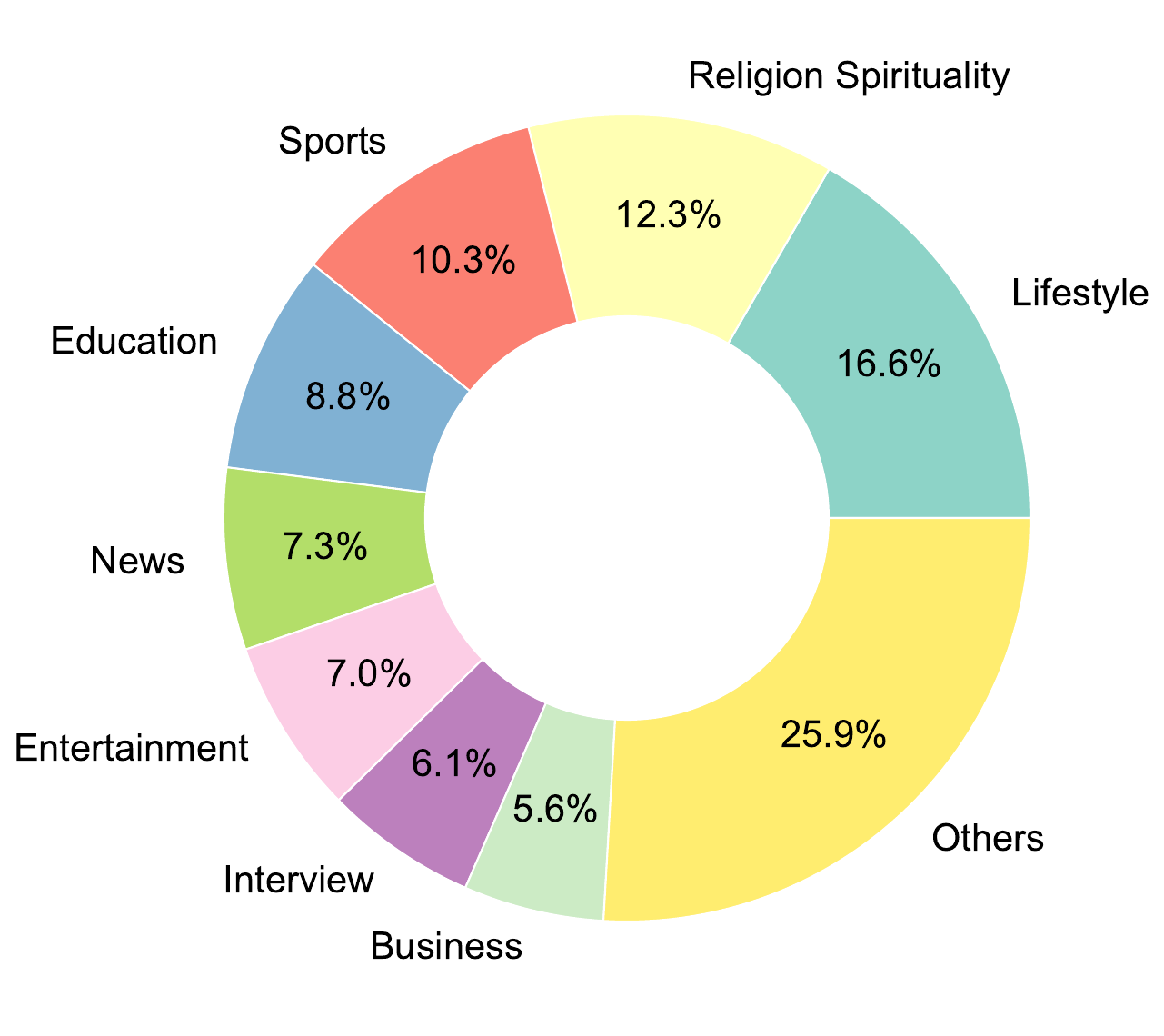}
    \caption{Category-wise statistics of the dataset used for Moshi fine-tuning experiments.}
    \label{fig:finetuning_dataset_stats}
\end{figure*}

This section provides statistics for the data fine-tuned in Section~\ref{exp_finetuning}. Figure~\ref{fig:finetuning_dataset_stats} illustrates that our training data originates from a wide range of conversational domains.

\subsection{Full-Duplex-Bench 1.0: Metric Definitions}
\label{subsec:full_duplex_bench_metrics}

Full-Duplex-Bench 1.0 evaluates spoken dialogue models under full-duplex conditions, focusing on pause handling, backchanneling, smooth turn-taking, and user interruption handling. Across all tasks, we define latency for an instance $i$ as $\Delta_i = t_{\text{start},i} - t_{\text{end},i}$, where $t_{\text{start}}$ denotes the model's response onset and $t_{\text{end}}$ denotes the end of the relevant user event.

\paragraph{Pause Handling.}
To evaluate whether the model incorrectly treats mid-utterance silence as a turn boundary, we use \textbf{Synthetic/Candor TOR} (Turn-Over Rate, $\downarrow$). This metric calculates the fraction of pause instances in which the model starts speaking. A failure is recorded if the model's output during a pause exceeds a minimal threshold (duration $\ge 1$ second or $> 3$ words).

\paragraph{Backchanneling.}
We assess the model's ability to provide brief acknowledgements without seizing the floor using three metrics. \textbf{Backchannel TOR} ($\downarrow$) measures the fraction of backchannel-eligible windows where the model produces a full turn (duration $\ge 3$s or $\ge 1$s with $>3$ words). \textbf{Frequency} ($\uparrow$) reports the number of backchannels normalized by total audio duration. Finally, \textbf{JSD} ($\downarrow$) computes the Jensen--Shannon divergence between the model's backchannel timing distribution and human ground truth to evaluate timing naturalness.

\paragraph{Smooth Turn Taking.}
This task measures the model's promptness in responding after the user completes an utterance. We report \textbf{Candor TOR} ($\uparrow$), defined as the fraction of user-turn endings where the model successfully begins speaking, and \textbf{Latency} ($\downarrow$), measured only on instances where the model successfully takes the turn.

\paragraph{User Interruption.}
When a user interrupts the model, we evaluate the system's responsiveness and contextual adaptation. \textbf{Interruption TOR} ($\uparrow$) measures the fraction of interruption events where the model responds. \textbf{Latency} ($\downarrow$) tracks the time from the end of the interruption to the model's response. Additionally, we use a \textbf{GPT-4o relevance score} ($\uparrow$) (0--5 scale) to assess whether the model's response is semantically relevant to the content of the interruption.

\begin{table*}[t]
\centering
\caption{Full-Duplex-Bench v1.5 Evaluation Results (Moshi vs. Fine-tuned Moshi).
Comparison of post-distractor audio quality metrics and behavior classification across four overlap scenarios.}
\resizebox{\textwidth}{!}{%
\begin{tabular}{l|cccc|c|cc|cc|cccc}
\toprule
\rowcolor{naverbg}
\multicolumn{14}{c}{\opthead{Full-Duplex-Bench v1.5: Overlap Handling Evaluation}} \\
\midrule
\multirow{2}{*}{\opthead{Scenario}}
& \multicolumn{4}{c|}{\opthead{Audio Quality (Post)}}
& \opthead{Rate}
& \multicolumn{2}{c|}{\opthead{Pitch}}
& \multicolumn{2}{c|}{\opthead{Intensity}}
& \multicolumn{4}{c}{\opthead{Behavior Ratio}} \\
\cmidrule(lr){2-5} \cmidrule(lr){6-6} \cmidrule(lr){7-8} \cmidrule(lr){9-10} \cmidrule(lr){11-14}
& STOI$\uparrow$ & PESQ$\uparrow$ & SI-SDR$\uparrow$ & UTMOS$\uparrow$
& WPM
& $\mu$ & $\sigma$
& $\mu$ & $\sigma$
& RESP & RESU & UNCERT & UNK \\
\midrule
\multicolumn{14}{c}{\textit{\textbf{Moshi (Base)}}} \\
\midrule
Background Speech   & 0.79 & 2.19 & 5.43 & 1.86 & 75.9  & 85.9  & 11.5 & -64.6 & 16.3 & 0.15 & 0.07 & 0.03 & 0.75 \\
Talking to Other    & 0.90 & 2.55 & 12.64 & \textbf{2.34} & 124.5 & 96.6  & 16.0 & -49.1 & 16.2 & 0.15 & 0.18 & 0.04 & 0.63 \\
User Backchannel    & 0.63 & 1.60 & -6.57 & 1.25 & 25.8  & 66.5  & 6.2  & -87.0 & 16.0 & 0.01 & 0.06 & 0.01 & 0.92 \\
User Interruption   & 0.94 & 2.87 & 16.07 & \textbf{2.65} & 149.0 & 111.1 & 21.8 & -41.3 & 16.0 & 0.59 & 0.17 & 0.03 & 0.21 \\
\midrule
\rowcolor{naverbg}\multicolumn{14}{c}{\textit{\textbf{Moshi (Fine-tuned)}}} \\
\midrule
Background Speech   & \textbf{0.98} & \textbf{3.33} & \textbf{20.76} & \textbf{1.87} & 157.5 & 88.3  & 13.7 & -64.8 & 16.4 & 0.28 & 0.11 & 0.00 & 0.60 \\
Talking to Other    & \textbf{0.96} & \textbf{3.30} & \textbf{20.26} & 2.30 & 146.4 & 96.4  & 17.0 & -51.0 & 16.3 & 0.18 & 0.16 & 0.00 & 0.63 \\
User Backchannel    & \textbf{0.91} & \textbf{3.01} & \textbf{16.48} & \textbf{1.32} & 132.5 & 72.0  & 7.9  & -85.2 & 16.0 & 0.08 & 0.11 & 0.00 & 0.72 \\
User Interruption   & \textbf{0.97} & \textbf{3.27} & \textbf{20.26} & 2.58 & 156.0 & 110.7 & 22.8 & -43.8 & 15.7 & 0.51 & 0.12 & 0.00 & 0.36 \\

\bottomrule
\end{tabular}}
\label{tab:v1.5-results}
\end{table*}

\begin{table}[t]
\centering
\caption{Full-Duplex-Bench v1.5 Latency Analysis (Moshi vs. Fine-tuned Moshi).
Stop latency measures time from user speech onset to model speech cessation.
Response latency measures time from user speech offset to model speech resumption.}
\resizebox{0.7\textwidth}{!}{%
\begin{tabular}{l|cc|cc|cc}
\toprule
\rowcolor{naverbg}
\multicolumn{7}{c}{\opthead{Latency Analysis (seconds)}} \\
\midrule
\multirow{2}{*}{\opthead{Scenario}}
& \multicolumn{2}{c|}{\opthead{Stop Latency$\downarrow$}}
& \multicolumn{2}{c|}{\opthead{Response Latency$\downarrow$}}
& \multicolumn{2}{c}{\opthead{Sample Count}} \\
\cmidrule(lr){2-3} \cmidrule(lr){4-5} \cmidrule(lr){6-7}
& $\mu$ & $\sigma$
& $\mu$ & $\sigma$
& Stop & Resp \\
\midrule
\multicolumn{7}{c}{\textit{\textbf{Moshi (Base)}}} \\
\midrule
Background Speech   & 1.02 & 0.55 & 2.90 & 2.05 & 150 & 89  \\
Talking to Other    & 1.13 & \textbf{0.57} & 3.22 & 1.87 & 184 & 117 \\
User Backchannel    & 1.30 & 0.42 & 2.38 & 1.84 & 113 & 29  \\
User Interruption   & 1.30 & 0.72 & 1.99 & 2.24 & 391 & 237 \\
\midrule
\rowcolor{naverbg}\multicolumn{7}{c}{\textit{\textbf{Moshi (Fine-tuned)}}} \\
\midrule
Background Speech   & \textbf{0.68} & \textbf{0.48} & \textbf{0.73} & \textbf{0.49} & 44  & 192 \\
Talking to Other    & \textbf{0.82} & 0.65 & \textbf{0.84 }& \textbf{0.69} & 47  & 188 \\
User Backchannel    & \textbf{0.70} & \textbf{0.40} & \textbf{1.12} & \textbf{0.85} & 57  & 156 \\
User Interruption   & \textbf{0.89} & \textbf{0.64} & \textbf{0.66} & \textbf{0.55} & 110 & 383 \\
\bottomrule
\end{tabular}}
\label{tab:v1.5-latency}
\end{table}

\subsection{Results on Full-Duplex-Bench 1.5} 
\label{results_on_full-dup_1.5}
\citet{fullduplexbench2} released Full-Duplex-Bench 1.5 as a successor to v1.0. We further evaluated the model fine-tuned on \textsc{Sommelier}-processed data using this updated benchmark. 

The experimental results presented in Table~\ref{tab:v1.5-results} and Table~\ref{tab:v1.5-latency} demonstrate that the \textsc{Sommelier}-fine-tuned Moshi model significantly outperforms the base model across all overlap scenarios.
In terms of audio quality, the fine-tuned model exhibits superior signal fidelity and robustness, evidenced by substantial gains in PESQ and SI-SDR scores; notably, the SI-SDR for the `Background Speech' scenario improved drastically from 5.43~dB to 20.76~dB.
Furthermore, the latency analysis reveals a critical enhancement in conversational responsiveness, with both stop and response latencies reduced to sub-second averages in the majority of cases, thereby enabling more natural and immediate turn-taking interactions.

\section{Context Captioning}
\label{context_captioning}
Speech data contains rich non-verbal information, such as timbre and emotion, beyond text semantics~\citep{koolagudi2012emotion, tursunov2019discriminating, lee2025dittottsdiffusiontransformersscalable, TeXBLEU}.
Detailed captioning of this information serves as effective metadata for speech understanding and generation~\citep{wang2025capspeech, ai2025ming, yan2025minguniaudiospeechllmjoint}.
Unlike other studies, we propose captioning audio segments using the Qwen3-Omni-Captioner~\citep{xu2025qwen3omnitechnicalreport} model to generate rich metadata, including emotion, gender, age group, and situation descriptions.

However, captioning short segments individually can fail to capture context (e.g., sarcasm). To address this, we implemented context-aware captioning by providing the preceding two segments as audio prompts (In-Context Learning). Specifically, for consecutive audio segments $a_1$, $a_2$, and $a_3$, we calculate the conditional probability $P(C_3|I, a_1, a_2)$ to generate the caption $C_3$ for $a_3$.

%\newpage
\section{Example}
\label{example}
\begin{figure*}[h]
    \centering
    % TODO: replace with your desired image file
    \includegraphics[width=1\linewidth]{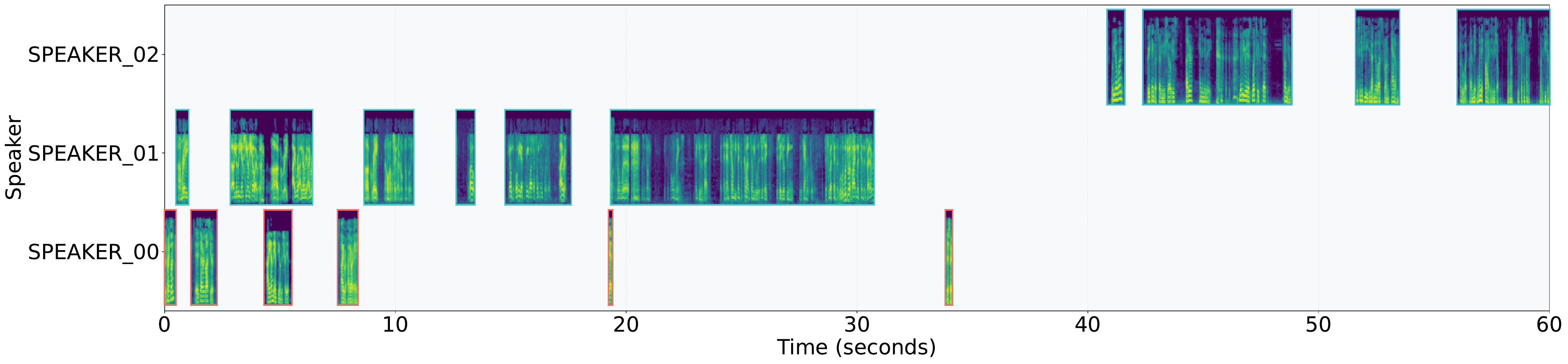}
    \caption{Visualization of preprocessing results for a 1-minute audio clip using a mel-spectrogram.}
    \label{fig:appendix-example-image}
\end{figure*}

This section presents real-world podcast examples. Figure~\ref{fig:appendix-example-image} visualizes the data processed by \textsc{Sommelier}, followed by an example of the corresponding JSON file.

\label{app:json_example}
\newpage

\begin{lstlisting}[
    language=json,
    numbers=left,    
    numbersep=5pt,    
    xleftmargin=2em,   
    frame=single,      
    breaklines=true     
]
{
  "metadata": {
    "audio_duration_seconds": 120.0,
    "audio_duration_minutes": 2.0,
    "vad_sortformer": {
      "processing_time_seconds": 0.9742708206176758,
      "rt_factor": 0.008118923505147297
    },
    "whisper_large_v3": {
      "processing_time_seconds": 14.903292655944824,
      "rt_factor": 0.12419410546620686
    },
    "total_segments": 26,
    "whisperx_alignment": {
      "processing_time_seconds": 32.398998737335205,
      "rt_factor": 0.26999165614446,
      "enabled": true
    },
    "sepreformer_separation": {
      "processing_time_seconds": 0.12217402458190918,
      "rt_factor": 0.00101811687151591,
      "overlap_threshold_seconds": 0.2,
      "enabled": true
    },
    "flowse_denoising": {
      "processing_time_seconds": 3.8639004230499268,
      "rt_factor": 0.032199170192082724,
      "enabled": true
    }
  },
  "segments": [
    {
      "start": 0.0,
      "end": 0.64,
      "text": "Mr. Franklin?",
      "text_whisper": "Mr. Franklin?",
      "text_parakeet": "The Franklin?",
      "text_canary": "The Franklin",
      "speaker": "SPEAKER_00",
      "language": "en",
      "demucs": false,
      "is_separated": true,
      "sepreformer": false,
      "words": [
        {
          "word": "Mr.",
          "start": 0.0,
          "end": 0.171,
          "score": 0.414
        },
        {
          "word": "Franklin?",
          "start": 0.192,
          "end": 0.661,
          "score": 0.936
        }
      ]
    },
    {
      "start": 0.48,
      "end": 1.2,
      "text": "I'm ready.",
      "text_whisper": "I'm ready.",
      "text_parakeet": "I'm ready.",
      "text_canary": "I'm ready",
      "speaker": "SPEAKER_01",
      "language": "en",
      "demucs": false,
      "is_separated": true,
      "sepreformer": false,
      "words": [
        {
          "word": "I'm",
          "start": 0.48,
          "end": 0.861,
          "score": 0.611
        },
        {
          "word": "ready.",
          "start": 0.9039999999999999,
          "end": 1.221,
          "score": 0.748
        }
      ]
    },
    {
      "start": 1.12,
      "end": 2.5599999999999996,
      "text": "It's Ira Glass here",
      "text_whisper": "Tyra Glass here.",
      "text_parakeet": "It's Iraq Glass here.",
      "text_canary": "It's Ira Glass here",
      "speaker": "SPEAKER_00",
      "language": "en",
      "demucs": false,
      "is_separated": true,
      "sepreformer": false,
      "words": [
        {
          "word": "Tyra",
          "start": 1.12,
          "end": 1.9220000000000002,
          "score": 0.686
        },
        {
          "word": "Glass",
          "start": 1.943,
          "end": 2.1900000000000004,
          "score": 0.801
        },
        {
          "word": "here.",
          "start": 2.21,
          "end": 2.5810000000000004,
          "score": 0.75
        }
      ]
    },
    {
      "start": 2.8,
      "end": 7.279999999999999,
      "text": "Oh you're the MC on the show I read about Oh great I read I read I read",
      "text_whisper": "You're the emcee on the show, Ira. Oh, great. Ira, are you Ira? Ira?",
      "text_parakeet": "Oh, you're the MC on the show, Ira. Oh, great. Ira Iron.",
      "text_canary": "Oh you're the MC on the show I read about Oh great I read I read I read",
      "speaker": "SPEAKER_01",
      "language": "en",
      "demucs": false,
      "is_separated": true,
      "sepreformer": true,
      "words": [
        {
          "word": "You're",
          "start": 2.8,
          "end": 3.2439999999999998,
          "score": 0.498
        },
        {
          "word": "the",
          "start": 3.264,
          "end": 3.3649999999999998,
          "score": 0.685
        },
        {
          "word": "emcee",
          "start": 3.4459999999999997,
          "end": 3.7479999999999998,
          "score": 0.678
        },
        {
          "word": "on",
          "start": 3.7889999999999997,
          "end": 3.8489999999999998,
          "score": 0.932
        },
        {
          "word": "the",
          "start": 3.87,
          "end": 3.9299999999999997,
          "score": 0.977
        },
        {
          "word": "show,",
          "start": 3.9499999999999997,
          "end": 4.213,
          "score": 0.72
        },
        {
          "word": "Ira.",
          "start": 4.233,
          "end": 5.282,
          "score": 0.799
        },
        {
          "word": "Oh,",
          "start": 5.302,
          "end": 5.484,
          "score": 0.832
        },
        {
          "word": "great.",
          "start": 5.645,
          "end": 5.968,
          "score": 0.85
        },
        {
          "word": "Ira,",
          "start": 6.311,
          "end": 6.614,
          "score": 0.534
        },
        {
          "word": "are",
          "start": 6.634,
          "end": 6.695,
          "score": 0.211
        },
        {
          "word": "you",
          "start": 6.775,
          "end": 6.936999999999999,
          "score": 0.317
        },
        {
          "word": "Ira?",
          "start": 6.957,
          "end": 7.119,
          "score": 0.531
        },
        {
          "word": "Ira?",
          "start": 7.139,
          "end": 7.3,
          "score": 0.582
        }
      ],
      "flowse_denoised": true
    },
    {
      "start": 4.24,
      "end": 5.76,
      "text": "I am the MC on this show, yes.",
      "text_whisper": "I am the MC on this show, yes.",
      "text_parakeet": "I am the MC on the show, yes.",
      "text_canary": "I am the MC on this show yes",
      "speaker": "SPEAKER_00",
      "language": "en",
      "demucs": false,
      "is_separated": true,
      "sepreformer": true,
      "words": [
        {
          "word": "I",
          "start": 4.24,
          "end": 4.61,
          "score": 0.838
        },
        {
          "word": "am",
          "start": 4.63,
          "end": 4.7330000000000005,
          "score": 0.946
        },
        {
          "word": "the",
          "start": 4.774,
          "end": 4.877000000000001,
          "score": 0.215
        },
        {
          "word": "MC",
          "start": 4.897,
          "end": 5.123,
          "score": 0.683
        },
        {
          "word": "on",
          "start": 5.144,
          "end": 5.205,
          "score": 0.882
        },
        {
          "word": "this",
          "start": 5.226,
          "end": 5.329000000000001,
          "score": 0.287
        },
        {
          "word": "show,",
          "start": 5.349,
          "end": 5.514,
          "score": 0.455
        },
        {
          "word": "yes.",
          "start": 5.534000000000001,
          "end": 5.781000000000001,
          "score": 0.818
        }
      ],
      "flowse_denoised": true
    },
    {
      "start": 7.36,
      "end": 8.48,
      "text": "IRA. IRA.",
      "text_whisper": "IRA. IRA.",
      "text_parakeet": "Ira, IRA.",
      "text_canary": "IRA I-R-A",
      "speaker": "SPEAKER_00",
      "language": "en",
      "demucs": false,
      "is_separated": true,
      "sepreformer": false,
      "words": [
        {
          "word": "IRA.",
          "start": 7.36,
          "end": 8.003,
          "score": 0.779
        },
        {
          "word": "IRA.",
          "start": 8.024000000000001,
          "end": 8.501000000000001,
          "score": 0.648
        }
      ]
    },
    {
      "start": 8.48,
      "end": 11.2,
      "text": "Oh, great. Now hold on one second Larry. Don't don't go away.",
      "text_whisper": "Oh, great. Now, hold on one second there. Don't go away.",
      "text_parakeet": "Oh, great. Now hold on one second Larry. Don't don't go away.",
      "text_canary": "Oh great now hold on one second there don't go away",
      "speaker": "SPEAKER_01",
      "language": "en",
      "demucs": false,
      "is_separated": true,
      "sepreformer": false,
      "words": [
        {
          "word": "Oh,",
          "start": 8.48,
          "end": 8.825000000000001,
          "score": 0.766
        },
        {
          "word": "great.",
          "start": 8.906,
          "end": 9.292,
          "score": 0.951
        },
        {
          "word": "Now,",
          "start": 9.678,
          "end": 9.759,
          "score": 0.146
        },
        {
          "word": "hold",
          "start": 9.779,
          "end": 9.881,
          "score": 0.504
        },
        {
          "word": "on",
          "start": 9.941,
          "end": 10.002,
          "score": 0.515
        },
        {
          "word": "one",
          "start": 10.063,
          "end": 10.144,
          "score": 0.348
        },
        {
          "word": "second",
          "start": 10.165000000000001,
          "end": 10.327,
          "score": 0.295
        },
        {
          "word": "there.",
          "start": 10.347000000000001,
          "end": 10.693000000000001,
          "score": 0.322
        },
        {
          "word": "Don't",
          "start": 10.733,
          "end": 10.875,
          "score": 0.69
        },
        {
          "word": "go",
          "start": 10.896,
          "end": 10.997,
          "score": 0.406
        },
        {
          "word": "away.",
          "start": 11.017,
          "end": 11.22,
          "score": 0.513
        }
      ]
    },
    {
      "start": 12.4,
      "end": 13.44,
      "text": "Hello?",
      "text_whisper": "Hello?",
      "text_parakeet": "Hello?",
      "text_canary": "Hello",
      "speaker": "SPEAKER_01",
      "language": "en",
      "demucs": false,
      "is_separated": true,
      "sepreformer": false,
      "words": [
        {
          "word": "Hello?",
          "start": 12.4,
          "end": 13.461,
          "score": 0.415
        }
      ]
    },
    {
      "start": 14.48,
      "end": 18.080000000000002,
      "text": "Sheldon, call me after 3 o'clock. I've got great news for you. Ira...",
      "text_whisper": "Sheldon, call me after 3 o'clock. I've got great news for you. Ira...",
      "text_parakeet": "Sheldon, call me after three o'clock. I've got great news for you. Ira.",
      "text_canary": "Shelton McCoomy at three o'clock got great news for you Irum",
      "speaker": "SPEAKER_01",
      "language": "en",
      "demucs": false,
      "is_separated": true,
      "sepreformer": false,
      "words": [
        {
          "word": "Sheldon,",
          "start": 14.48,
          "end": 15.046000000000001,
          "score": 0.336
        },
        {
          "word": "call",
          "start": 15.067,
          "end": 15.208,
          "score": 0.237
        },
        {
          "word": "me",
          "start": 15.228,
          "end": 15.329,
          "score": 0.714
        },
        {
          "word": "after",
          "start": 15.39,
          "end": 15.552,
          "score": 0.484
        },
        {
          "word": "3",
          "start": 15.572000000000001,
          "end": 15.754000000000001,
          "score": 0.444
        },
        {
          "word": "o'clock.",
          "start": 15.774000000000001,
          "end": 16.017,
          "score": 0.782
        },
        {
          "word": "I've",
          "start": 16.037,
          "end": 16.118000000000002,
          "score": 0.003
        },
        {
          "word": "got",
          "start": 16.138,
          "end": 16.199,
          "score": 0.242
        },
        {
          "word": "great",
          "start": 16.219,
          "end": 16.401,
          "score": 0.784
        },
        {
          "word": "news",
          "start": 16.442,
          "end": 16.624000000000002,
          "score": 0.566
        },
        {
          "word": "for",
          "start": 16.664,
          "end": 16.806,
          "score": 0.624
        },
        {
          "word": "you.",
          "start": 16.846,
          "end": 17.008,
          "score": 0.733
        },
        {
          "word": "Ira...",
          "start": 17.615000000000002,
          "end": 18.1,
          "score": 0.818
        }
      ]
    },
    {
      "start": 18.88,
      "end": 19.12,
      "text": "Yes.",
      "text_whisper": "Yes.",
      "text_parakeet": "Yeah.",
      "text_canary": "Yeah",
      "speaker": "SPEAKER_00",
      "language": "en",
      "demucs": false,
      "is_separated": true,
      "sepreformer": false,
      "words": [
        {
          "word": "Yes.",
          "start": 18.88,
          "end": 19.144,
          "score": 0.471
        }
      ]
    },
    {
      "start": 18.96,
      "end": 33.28,
      "text": "So uh listen, Tony if the phone rings, take it in the back and then tell me then come out and tell me who it is. is. Just Just say Joe's being with a camera crew. Just for about 10 minutes. We'll do about five minutes, ten minutes, right, Irv?",
      "text_whisper": "So, listen, Tony, if the phone rings, take it in the back, and then come out and tell me who it is. Just say Joe's being with a camera crew. Just for about ten minutes. We'll do about five minutes, ten minutes, right, Iris?",
      "text_parakeet": "So listen, Tony. If the phone rings, take it in the back and tell me, then come out and tell me who it is. Just say Joe's being with a camera crew. Just for about 10 minutes. We'll do about five minutes, ten minutes, right, Ivory?",
      "text_canary": "So uh listen Tony if the phone rings take it in the back and then tell me then come out and tell me who it is just say just say Joe's being with the camera crew just for about 10 minutes we'll do a five minute ten minutes right Irv?",
      "speaker": "SPEAKER_01",
      "language": "en",
      "demucs": true,
      "is_separated": true,
      "sepreformer": false,
      "words": [
        {
          "word": "So,",
          "start": 18.96,
          "end": 19.983,
          "score": 0.798
        },
        {
          "word": "listen,",
          "start": 20.685000000000002,
          "end": 20.905,
          "score": 0.823
        },
        {
          "word": "Tony,",
          "start": 20.946,
          "end": 21.186,
          "score": 0.799
        },
        {
          "word": "if",
          "start": 21.928,
          "end": 22.029,
          "score": 0.914
        },
        {
          "word": "the",
          "start": 22.069000000000003,
          "end": 22.169,
          "score": 0.872
        },
        {
          "word": "phone",
          "start": 22.229,
          "end": 22.51,
          "score": 0.66
        },
        {
          "word": "rings,",
          "start": 22.57,
          "end": 22.851,
          "score": 0.851
        },
        {
          "word": "take",
          "start": 23.573,
          "end": 23.753,
          "score": 0.923
        },
        {
          "word": "it",
          "start": 23.814,
          "end": 23.854,
          "score": 0.979
        },
        {
          "word": "in",
          "start": 23.894000000000002,
          "end": 23.934,
          "score": 0.989
        },
        {
          "word": "the",
          "start": 23.974,
          "end": 24.034,
          "score": 0.967
        },
        {
          "word": "back,",
          "start": 24.094,
          "end": 24.335,
          "score": 0.953
        },
        {
          "word": "and",
          "start": 25.157,
          "end": 25.438000000000002,
          "score": 0.708
        },
        {
          "word": "then",
          "start": 25.839,
          "end": 26.02,
          "score": 0.77
        },
        {
          "word": "come",
          "start": 26.28,
          "end": 26.381,
          "score": 0.906
        },
        {
          "word": "out",
          "start": 26.401,
          "end": 26.481,
          "score": 0.957
        },
        {
          "word": "and",
          "start": 26.521,
          "end": 26.581000000000003,
          "score": 0.993
        },
        {
          "word": "tell",
          "start": 26.601,
          "end": 26.722,
          "score": 0.854
        },
        {
          "word": "me",
          "start": 26.762,
          "end": 26.822000000000003,
          "score": 0.966
        },
        {
          "word": "who",
          "start": 26.862000000000002,
          "end": 26.962000000000003,
          "score": 0.963
        },
        {
          "word": "it",
          "start": 27.002000000000002,
          "end": 27.043,
          "score": 0.879
        },
        {
          "word": "is.",
          "start": 27.103,
          "end": 27.163,
          "score": 0.86
        },
        {
          "word": "Just",
          "start": 27.183,
          "end": 27.303,
          "score": 0.927
        },
        {
          "word": "say",
          "start": 27.343,
          "end": 27.444000000000003,
          "score": 0.782
        },
        {
          "word": "Joe's",
          "start": 28.306,
          "end": 28.527,
          "score": 0.608
        },
        {
          "word": "being",
          "start": 28.567,
          "end": 28.767000000000003,
          "score": 0.883
        },
        {
          "word": "with",
          "start": 29.229,
          "end": 29.349,
          "score": 0.822
        },
        {
          "word": "a",
          "start": 29.389000000000003,
          "end": 29.409,
          "score": 0.432
        },
        {
          "word": "camera",
          "start": 29.449,
          "end": 29.79,
          "score": 0.854
        },
        {
          "word": "crew.",
          "start": 29.810000000000002,
          "end": 30.011000000000003,
          "score": 0.585
        },
        {
          "word": "Just",
          "start": 30.633000000000003,
          "end": 30.773000000000003,
          "score": 0.973
        },
        {
          "word": "for",
          "start": 30.793,
          "end": 30.873,
          "score": 0.847
        },
        {
          "word": "about",
          "start": 30.893,
          "end": 31.034,
          "score": 0.98
        },
        {
          "word": "ten",
          "start": 31.074,
          "end": 31.234,
          "score": 0.669
        },
        {
          "word": "minutes.",
          "start": 31.254,
          "end": 31.535,
          "score": 0.69
        },
        {
          "word": "We'll",
          "start": 31.555,
          "end": 31.816000000000003,
          "score": 0.449
        },
        {
          "word": "do",
          "start": 31.836,
          "end": 31.956000000000003,
          "score": 0.81
        },
        {
          "word": "about",
          "start": 31.976,
          "end": 32.077,
          "score": 0.188
        },
        {
          "word": "five",
          "start": 32.117000000000004,
          "end": 32.317,
          "score": 0.854
        },
        {
          "word": "minutes,",
          "start": 32.357,
          "end": 32.538,
          "score": 0.359
        },
        {
          "word": "ten",
          "start": 32.578,
          "end": 32.698,
          "score": 0.858
        },
        {
          "word": "minutes,",
          "start": 32.718,
          "end": 32.919,
          "score": 0.925
        },
        {
          "word": "right,",
          "start": 32.939,
          "end": 33.079,
          "score": 0.728
        },
        {
          "word": "Iris?",
          "start": 33.099000000000004,
          "end": 33.3,
          "score": 0.277
        }
      ]
    },
    {
      "start": 33.2,
      "end": 33.6,
      "text": "That's right.",
      "text_whisper": "That's right.",
      "text_parakeet": "Yep.",
      "text_canary": "That's fair",
      "speaker": "SPEAKER_00",
      "language": "en",
      "demucs": true,
      "is_separated": true,
      "sepreformer": false,
      "words": [
        {
          "word": "That's",
          "start": 33.2,
          "end": 33.444,
          "score": 0.426
        },
        {
          "word": "right.",
          "start": 33.489000000000004,
          "end": 33.622,
          "score": 0.164
        }
      ]
    },
    {
      "start": 40.08,
      "end": 41.04,
      "text": "Well you know what?",
      "text_whisper": "Well, you know what?",
      "text_parakeet": "Well you know what?",
      "text_canary": "Well you know what",
      "speaker": "SPEAKER_02",
      "language": "en",
      "demucs": true,
      "is_separated": true,
      "sepreformer": false,
      "words": [
        {
          "word": "Well,",
          "start": 40.08,
          "end": 40.455999999999996,
          "score": 0.865
        },
        {
          "word": "you",
          "start": 40.477,
          "end": 40.539,
          "score": 0.823
        },
        {
          "word": "know",
          "start": 40.580999999999996,
          "end": 40.684999999999995,
          "score": 0.822
        },
        {
          "word": "what?",
          "start": 40.79,
          "end": 41.061,
          "score": 0.358
        }
      ]
    },
    {
      "start": 41.6,
      "end": 49.68,
      "text": "Great Great thing about starting a new show is utter anonymity. Nobody really knows what to expect from you.",
      "text_whisper": "Great thing about starting a new show is utter anonymity. Nobody really knows what to expect from you.",
      "text_parakeet": "Great thing about starting a new show is utter anonymity. Nobody who really knows what to expect from you.",
      "text_canary": "The great thing about starting a new show is utter anonymity. Nobody really knows what to expect from you.",
      "speaker": "SPEAKER_02",
      "language": "en",
      "demucs": true,
      "is_separated": true,
      "sepreformer": false,
      "words": [
        {
          "word": "Great",
          "start": 41.6,
          "end": 42.042,
          "score": 0.86
        },
        {
          "word": "thing",
          "start": 42.062000000000005,
          "end": 42.203,
          "score": 0.797
        },
        {
          "word": "about",
          "start": 42.243,
          "end": 42.384,
          "score": 0.898
        },
        {
          "word": "starting",
          "start": 42.424,
          "end": 42.685,
          "score": 0.591
        },
        {
          "word": "a",
          "start": 42.705,
          "end": 42.726,
          "score": 0.669
        },
        {
          "word": "new",
          "start": 42.766,
          "end": 42.866,
          "score": 0.486
        },
        {
          "word": "show",
          "start": 42.906,
          "end": 43.107,
          "score": 0.976
        },
        {
          "word": "is",
          "start": 43.128,
          "end": 43.369,
          "score": 0.715
        },
        {
          "word": "utter",
          "start": 43.409,
          "end": 44.233000000000004,
          "score": 0.718
        },
        {
          "word": "anonymity.",
          "start": 44.273,
          "end": 45.298,
          "score": 0.852
        },
        {
          "word": "Nobody",
          "start": 46.866,
          "end": 47.107,
          "score": 0.66
        },
        {
          "word": "really",
          "start": 47.147,
          "end": 47.268,
          "score": 0.245
        },
        {
          "word": "knows",
          "start": 47.288000000000004,
          "end": 47.469,
          "score": 0.63
        },
        {
          "word": "what",
          "start": 47.57,
          "end": 47.75,
          "score": 0.788
        },
        {
          "word": "to",
          "start": 47.771,
          "end": 47.871,
          "score": 0.827
        },
        {
          "word": "expect",
          "start": 47.931000000000004,
          "end": 48.313,
          "score": 0.832
        },
        {
          "word": "from",
          "start": 48.353,
          "end": 49.399,
          "score": 0.958
        },
        {
          "word": "you.",
          "start": 49.459,
          "end": 49.7,
          "score": 0.861
        }
      ]
    },
    {
      "start": 50.64,
      "end": 53.04,
      "text": "This interviewee did not know us from Adam.",
      "text_whisper": "This interviewee did not know us from Adam.",
      "text_parakeet": "This interviewee did not know us from Adam.",
      "text_canary": "This interviewee did not know us from Adam",
      "speaker": "SPEAKER_02",
      "language": "en",
      "demucs": true,
      "is_separated": true,
      "sepreformer": false,
      "words": [
        {
          "word": "This",
          "start": 50.64,
          "end": 50.864,
          "score": 0.856
        },
        {
          "word": "interviewee",
          "start": 50.904,
          "end": 51.352000000000004,
          "score": 0.837
        },
        {
          "word": "did",
          "start": 51.393,
          "end": 51.494,
          "score": 0.847
        },
        {
          "word": "not",
          "start": 51.535000000000004,
          "end": 51.637,
          "score": 0.989
        },
        {
          "word": "know",
          "start": 51.698,
          "end": 51.86,
          "score": 0.847
        },
        {
          "word": "us",
          "start": 51.982,
          "end": 52.064,
          "score": 0.98
        },
        {
          "word": "from",
          "start": 52.206,
          "end": 52.409,
          "score": 0.956
        },
        {
          "word": "Adam.",
          "start": 52.43,
          "end": 53.06,
          "score": 0.447
        }
      ]
    },
    {
      "start": 54.96,
      "end": 63.04,
      "text": "Okay, well what? About a minute. We're one minute five into the new show right now. it is stretching in front of us. The perfect future.",
      "text_whisper": "Okay, well, what? About a minute. We're one minute five into the new show right now. It is stretching in front of us. The perfect future.",
      "text_parakeet": "Okay, well what? About a minute. We're one minute five into the new show. Right now, it is stretching in front of us. A perfect future.",
      "text_canary": "Okay well what about a minute well one minute five into the new show right now it is stretching in front of us a perfect future",
      "speaker": "SPEAKER_02",
      "language": "en",
      "demucs": true,
      "is_separated": true,
      "sepreformer": false,
      "words": [
        {
          "word": "Okay,",
          "start": 54.96,
          "end": 55.462,
          "score": 0.22
        },
        {
          "word": "well,",
          "start": 55.483000000000004,
          "end": 55.583,
          "score": 0.241
        },
        {
          "word": "what?",
          "start": 55.603,
          "end": 55.804,
          "score": 0.764
        },
        {
          "word": "About",
          "start": 56.086,
          "end": 56.246,
          "score": 0.999
        },
        {
          "word": "a",
          "start": 56.307,
          "end": 56.327,
          "score": 0.999
        },
        {
          "word": "minute.",
          "start": 56.407000000000004,
          "end": 56.648,
          "score": 0.943
        },
        {
          "word": "We're",
          "start": 56.668,
          "end": 56.829,
          "score": 0.377
        },
        {
          "word": "one",
          "start": 56.89,
          "end": 56.97,
          "score": 0.891
        },
        {
          "word": "minute",
          "start": 57.01,
          "end": 57.211,
          "score": 0.904
        },
        {
          "word": "five",
          "start": 57.251,
          "end": 57.774,
          "score": 0.805
        },
        {
          "word": "into",
          "start": 57.814,
          "end": 58.055,
          "score": 0.774
        },
        {
          "word": "the",
          "start": 58.075,
          "end": 58.156,
          "score": 0.828
        },
        {
          "word": "new",
          "start": 58.196,
          "end": 58.317,
          "score": 0.597
        },
        {
          "word": "show",
          "start": 58.337,
          "end": 58.538000000000004,
          "score": 0.791
        },
        {
          "word": "right",
          "start": 58.578,
          "end": 59.502,
          "score": 0.883
        },
        {
          "word": "now.",
          "start": 59.543,
          "end": 59.724000000000004,
          "score": 0.982
        }
      ]
    }
  ]
}
\end{lstlisting}

\end{document}